\documentclass[10pt]{bmc_article}

\usepackage{cite} 
\usepackage{url}  
\usepackage{ifthen}  
\usepackage{multicol}   
\usepackage[utf8]{inputenc} 
\usepackage{bm}
\usepackage{graphicx}
\usepackage{amsfonts}
\usepackage{amssymb}
\usepackage{amsmath}
\usepackage{bbm}
\usepackage{latexsym}
\newboolean{publ}

\newenvironment{bmcformat}{\begin{raggedright}\baselineskip20pt\sloppy\setboolean{publ}{false}}{\end{raggedright}\baselineskip20pt\sloppy}

\newtheorem{theorem}{Theorem}
\newcommand{\bth}{\begin{theorem}}
\newcommand{\enth}{\end{theorem}}

\newtheorem{proposition}{Proposition}
\newcommand{\bp}{\begin{proposition}}
\newcommand{\ep}{\end{proposition}}

\newtheorem{corrollary}{Corrollary}
\newcommand{\bcor}{\begin{corrollary}}
\newcommand{\ecor}{\end{corrollary}}
\newtheorem{lemma}{Lemma}
\newcommand{\blem}{\begin{lemma}}
\newcommand{\elem}{\end{lemma}}
\newcommand{\bpb}{\begin{question}}
\newcommand{\epb}{\end{question}}

\newtheorem{definition}{Definition}
\newcommand{\bdf}{\begin{definition}}
\newcommand{\edf}{\end{definition}}
\newcommand{\bpr} {\noindent {\bf Proof }}
\newcommand{\epr}{\rule{0.25cm}{0.25cm}\\}

\newcommand\bbbr{{\sf I\!R}}

\newcommand\bbbz{{\sf I\!Z}}

\newcommand{\bd}{\begin{document}}
\newcommand{\ed}{\end{document}}
\newcommand{\beq}{\begin{equation}}
\newcommand{\eeq}{\end{equation}}
\newcommand{\bef}{\begin{figure}}
\newcommand{\enf}{\end{figure}}
\newcommand{\bea}{\begin{eqnarray}}
\newcommand{\eea}{\end{eqnarray}}
\newcommand{\baR}{\begin{array}}
\newcommand{\eaR}{\end{array}}
\newcommand{\bc}{\begin{center}}
\newcommand{\ec}{\end{center}}
\newcommand{\ben}{\begin{enumerate}}
\newcommand{\een}{\end{enumerate}}
\newcommand{\bit}{\begin{itemize}}
\newcommand{\eit}{\end{itemize}}
\newcommand{\su}{\section}
\newcommand{\ssu}{\subsection}
\newcommand{\sssu}{\subsubsection}
\newcommand{\nid}{\noindent}
\newcommand{\nnb}{\nonumber}

\newcommand\be{{\bf e}}
\newcommand\bv{{\bf v}}
\newcommand\h{{\bf h}}
\newcommand\bom{\mbox{{\boldmath $\omega$}}}
\newcommand\tom{\omega}
\newcommand\cM{{\cal M}}
\newcommand\cE{{\cal E}}
\newcommand\hM{{\hat{\cal M}}}
\newcommand\Mom{\cM_{\omega}}
\newcommand\on{\seq{\omega}{0}{n}}
\newcommand\onm{\seq{\omega}{0}{n-1}}
\newcommand\onmd{\left[\omega\right]_{n-2}}
\newcommand\ot{\left[\omega\right]_t}
\newcommand\otm{\left[\omega\right]_{t-1}}
\newcommand\Mon{\cM_{\on}}
\newcommand\Monm{\cM_{\onm}}
\newcommand\hMon{\hat{\cM}_{\on}}
\newcommand\hMot{\hat{\cM}_{\ot}}
\newcommand\oMon{\stackrel{\circ}{\cM}_{\on}}
\newcommand\ohMon{\stackrel{\circ}{\hM}_{\on}}
\newcommand\Anion{A_{i,n,\on}}
\newcommand\bAnion{\bar{A}_{i,n,\on}}
\newcommand\Fon{\F_{\on}}
\newcommand\Fot{\F_{\ot}}
\newcommand\hF{\hat{\F}}
\newcommand\hFon{\hF_{\on}}
\newcommand\hFot{\hF_{\ot}}
\newcommand\V{{\bf V}}
\newcommand\sV{\bf{\tiny{V}}}
\newcommand\f{{\bf{f}}}
\newcommand\hV{{\hat{\V}}}
\newcommand\hS{{\hat{\cS}}}
\newcommand\hml{{\hat{\mu}_0}}
\newcommand\piu{\pi^{(u)}}
\newcommand\pis{\pi^{(s)}}
\newcommand\tin{\tau_i(\omega,n)}

\newcommand\cEta{\cC(\bEta)}

\newcommand\bEtap{\mbox{{\boldmath $\eta'$}}}
\newcommand\cEtap{\cC(\bEtap)}
\newcommand\betat{\tilde{\bEta}}
\newcommand\betatp{\tilde{\bEtap}}
\newcommand\betatpl{\tilde{\bEta}^+}
\newcommand\bbeta{\mbox{{\boldmath $\beta$}}}
\newcommand\mub{\mu_{\bbeta}}
\newcommand\mus{\mu^\ast}
\newcommand\Ps{P^\ast}
\newcommand\Fb{\cF_{\bbeta}}
\newcommand\bxi{\bm{\xi}}
\newcommand\balpha{{\bm{\alpha}}}
\newcommand\blambda{{\bm{\lambda}}}
\newcommand\bphi{{\bm{\phi}}}
\newcommand\B{{\bf B}}
\newcommand\Vp{{\bf V'}}
\newcommand\W{{\bf W}}
\newcommand\Z{{\bf Z}}
\newcommand\F{{\bf F}}
\newcommand\Fe{\F_{\bEta}}
\newcommand\Fei{F_{\bEta,i}}
\newcommand\G{{\bf G}}
\newcommand\I{{\bf I}}
\newcommand\bz{{\bf 0}}
\newcommand\bzs{\left\{\bz\right\}}
\renewcommand\S{{\bf S}}
\newcommand\Td{T\left(\dAS\right)}
\newcommand\X{{\bf X}}
\newcommand\x{{\bf x}}
\newcommand\y{{\bf y}}
\newcommand\xt{\X_t(\V)}
\newcommand\cktv{\chi_{k,t}(\V)}
\newcommand\cA{{\cal A}}
\newcommand\cB{{\cal B}}
\newcommand\cH{{\cal H}}
\newcommand\cL{{\cal L}}
\newcommand\cN{{\cal N}}
\newcommand\cO{{\cal O}}
\newcommand\bme{\cB_\cM(\epsilon)}
\newcommand\Bev{\cB(\V,\epsilon)}
\newcommand\coMe{\stackrel{\circ}{\cMe}}
\newcommand\beV{\cB(\V,\epsilon)}
\newcommand\bmd{\cB_\cM(\delta)}
\newcommand\bmdp{\cB_\cM(\delta')}
\newcommand\cC{{\cal C}}
\newcommand\cD{{\cal D}}
\newcommand\cJ{{\cal J}}
\newcommand\cR{{\cal R}}
\newcommand\De{\cD(\bEta)}
\newcommand\bDe{\bar{\cD}(\bEta)}
\newcommand\Dep{\cD(\bEtap)}
\newcommand\bDep{\bar{\cD}(\bEtap)}
\newcommand\cG{{\cal G}}
\newcommand\GW{\cG_{(\cW,\Ie)}}
\newcommand\IW{\cI_{(\cW,\Ie)}}
\newcommand\cF{{\cal F}}
\newcommand\bcF{\bar{\cF}}
\newcommand\oGF{{\omega(\GF)}}
\newcommand\oGFTe{{\omega(\GFTe)}}
\newcommand\oF{{\omega_{\cF}(\GF)}}
\newcommand\oFTe{{\omega_{\cF}(\GFTe)}}
\newcommand\Bie{\cB_{\infty,\epsilon}}
\newcommand\BTe{\cB_{T,\epsilon}}
\newcommand\BTep{\cB_{{T+1},\epsilon}}
\newcommand\BTeu{\cB_{T,\epsilon}^{(1)}}
\newcommand\BTed{\cB_{T,\epsilon}^{(2)}}
\newcommand\DFFTV{\left. D\FTF\right|_{\V}}
\newcommand\FF{{\F_{\cF}}}
\newcommand\FTF{{\FT_{\cF}}}
\newcommand\FFb{{\F_{\bcF}}}
\newcommand\VF{{\V_{\cF}}}
\newcommand\VFb{{\V_{\bcF}}}
\newcommand\ft{\cF_t(\V)}
\newcommand\ftmu{\cF_{t-1}(\V)}
\newcommand\ftmk{\cF_{t-k}(\V)}
\newcommand\ftmkpu{\cF_{t-k+1}(\V)}
\newcommand\bft{\bar{\cF}_t(\V)}
\newcommand\bftmu{\bar{\cF}_{t-1}(\V)}
\newcommand\bftmk{\bar{\cF}_{t-k}(\V)}
\newcommand\bftmkpu{\bar{\cF}_{t-k+1}(\V)}
\newcommand\cMz{{\cal M_{\bf{0}}}}
\newcommand\cMu{{\cal M_{\bf{1}}}}
\newcommand\cMe{{\cM_{\bEta}}}
\newcommand\cMeu{{\cM_{{\bEta}_1}}}
\newcommand\cMet{{\cM_{{\bEta}_t}}}
\newcommand\cMeT{{\cM_{{\bEta}_T}}}
\newcommand\cMep{{\cM_{\bEtap}}}
\newcommand\cU{{\cal U}}
\newcommand\cS{{\cal S}}
\newcommand\cI{{\cal I}}
\newcommand\cP{{\cal P}}
\newcommand\cQ{{\cal Q}}
\newcommand\cV{{\cal V}}
\newcommand\cW{{\cal W}}
\newcommand\Vpi{V^+_i}
\newcommand\vp{\cV^+}

\newcommand\Is{\I^s}
\newcommand\IsP{\I^{s\ast}}
\newcommand\Ise{\I^s(\bEta)}
\newcommand\IE{\I(\bEta)}
\newcommand\Isv{\I^s(\V)}
\newcommand\Isi{I^s_i}
\newcommand\Isj{I^s_j}
\newcommand\Iie{I_i(\bEta)}
\newcommand\Iiep{I_i(\bEta')}
\newcommand\Ije{I_j(\bEta)}
\newcommand\Ijep{I_j(\bEta')}
\newcommand\Isie{\Isi(\bEta)}
\newcommand\Isje{\Isj(\bEta)}
\newcommand\Isjep{\Isj(\bEtap)}
\newcommand\Isiv{\Isi(\V)}
\newcommand\Isvt{I^s_i(\V(t))}
\newcommand\Ie{{\bf I^{ext}}}
\newcommand\Iei{I_i}
\newcommand\Iej{I^{ext}_j}
\newcommand\tk{\tau^{(k)}}
\newcommand\tkp{\tau^{(k+1)}}
\newcommand\toi{\tau^{(1)}_i(\V)}
\newcommand\toik{\tau^{(k)}_i(\V)}
\newcommand\toike{\tau^{(k)}_i(\betat)}
\newcommand\toikpe{\tau^{(k+1)}_i(\V)}
\newcommand\toikep{\tau^{(k+1)}_i(\betat)}
\newcommand\dik{\delta_i^{(k)}(\V)}
\newcommand\PrW{\cP_\cW}
\newcommand\PrI{\cP_\Ie}
\newcommand\PrWi{\cP_{\cW,\Ie}}
\newcommand\PW{\cP_{(\cW,\Ie)}}
\newcommand\RW{\cR_{(\cW,\Ie)}}
\newcommand\cPT{\cP^{(T)}}
\newcommand\cMT{\cM_{{\bEta_0} \dots {\bEta_T}}}
\newcommand\FT{\F^T}
\newcommand\Ft{\F^t}
\newcommand\Ftp{\F^{t+1}}
\newcommand\Fttk{\F^{t_k}}
\newcommand\Fmu{\F^{-1}}
\newcommand\Fmd{\F^{-2}}
\newcommand\FmT{\F^{-T}}
\newcommand\dAS{d(\oM,\cS)}
\newcommand\dSV{d(\tVp,\cS)}
\newcommand\trV{\left\{\V(t)\right\}_{t \geq 0}}
\newcommand\tV{\tilde{\V}}
\newcommand\tVp{\tilde{\V}^+}
\newcommand\tpi{t^{pre}_i}
\newcommand\tpj{t^{post}_j}
\newcommand\mW{m_{\cW}}
\newcommand\sW{\sigma_{\cW}}
\newcommand\mI{m_{\Ie}}
\newcommand\sI{\sigma_{\Ie}}
\newcommand\sw{\sigma_{(\cW,\Ie)}}
\newcommand\sB{\sigma_B^2}
\newcommand\sd{\sigma^2}
\newcommand\Cst{[\betat]_{s,t}}
\newcommand\defCst{\left\{\betat | \bEta_s=\as, \dots, \bEta_t = \at    \right\}}
\newcommand\as{\mbox{{\boldmath{$\alpha$}}}_s}
\newcommand\at{\mbox{{\boldmath{$\alpha$}}}_t}
\newcommand\moytpsi{\bar{\psi}_\cW}
\newcommand\sn{\left\{1 \dots N \right\}}
\newcommand\GF{\Gamma_\cF}
\newcommand\GFTe{\Gamma_{\cF,T,\epsilon}}
\newcommand\GFTep{\Gamma_{\cF,T,\epsilon'}}
\newcommand\GFTpe{\Gamma_{\cF,{T+1},\epsilon}}
\newcommand\PF{\Pi_{\cF}}
\newcommand\PbF{\Pi_{\bcF}}
\newcommand\oM{\Omega}

\newcommand\wsg{\cW^s(\V)}
\newcommand\wsl{\cW^s_{loc}(\V)}
\newcommand\wslf{\cW^s_{loc}(\F(\V))}
\newcommand\wse{\cW^s_{\epsilon}(\V)}
\newcommand\wset{\cW^s_{\epsilon}(\V(t))}
\newcommand\tik{\tau_i^{(k)}}
\newcommand\tiu{\tau_i^{(1)}}
\newcommand\tikp{\tau_i^{(k+1)}}
\newcommand\Vr{V_{reset}}
\newcommand\Vm{V_{min}}
\newcommand\VM{V_{max}}
\newcommand\Vs{V^{+}}
\newcommand\Vsi{V_i^{+}}
\newcommand\SLp{\Sigma_\Lambda^+}
\newcommand\SL{\Sigma_\Lambda}
\newcommand\SW{\Sigma_{(\cW,\Ie)}}
\newcommand\SWp{\Sigma_{(\cW,\Ie)}^+}
\newcommand{\D}{\displaystyle}
\newcommand{\deq}{\stackrel {\rm def}{=}}
\newcommand{\peq}{\stackrel {\rm \mu p.s.}{=}}

\newcommand\cyl[3]{\left[{#1}\right]_{#2}^{#3}}
\newcommand\seq[3]{{#1}_{#2}^{#3}}
\newcommand\bloc[2]{{\omega}_{#1}^{#2}}
\newcommand\somt{\seq{\omega}{-\infty}{t}}
\newcommand\prodset[3]{{#1}_{#2}^{#3}}
\newcommand\tinm{\tau_i(\seq{\omega}{-\infty}{n-1})}
\newcommand\uom{\underline\omega}
\newcommand\uX{\underline X}
\newcommand\ucF{\underline \cF}
\newcommand\uomp{\underline\omega'}
\newcommand\tiz{\tau_i(\uom)}
\newcommand{\setZ}{\mathbbm{Z}}
\newcommand{\setN}{\mathbbm{N}}
\newcommand{\setR}{\mathbbm{R}}

\newcommand{\ent}[1]{\left[\, #1 \, \right]}
\newcommand{\frc}[1]{\left\{\, #1 \, \right\}}
\newcommand{\abs}[1]{\left|\, #1 \, \right|}
\newcommand{\pare}[1]{\left(\, #1 \, \right)}
\newcommand{\bra}[1]{\left[\, #1 \, \right]}
\newcommand{\brac}[2]{\left[\, #1 \, | \, #2 \right]}
\newcommand{\Set}[1]{\left\{\, #1 \, \right\}}
\newcommand{\Setc}[2]{\left\{\, #1 \, ; \, #2 \right\}}
\newcommand{\Exp}[1]{\mathbb{E}\left[\, #1 \, \right]}
\newcommand{\Cov}[1]{Cov\left[\, #1 \, \right]}
\newcommand{\Expc}[2]{\mathbb{E}\left[\, #1 \, | \, #2 \right]}
\newcommand{\Covc}[2]{Cov\left[\, #1 \,\, | \, #2 \right]}
\newcommand{\Prob}[1]{P\left[\, #1 \, \right]}
\newcommand{\Probc}[2]{P\left[\, #1 \, \left| \, #2 \right. \right]}
\newcommand{\Pnc}[2]{P_n\left[\, #1 \, \left| \, #2 \right. \right]}
\newcommand{\Ptc}[2]{P_t\left[\, #1 \, \left| \, #2 \right. \right]}
\newcommand\pTo{{\pi_{\omega}^{(T)}}}
\newcommand\tj[2]{t_j^{(#1)}(#2)}
\newcommand\tjr[1]{\tj{r}{#1}}
\newcommand\tjro{\tjr{\omega}}
\newcommand\tjk{t_j^{(k)}(\omega)}
\newcommand\sif[1]{\seq{\omega}{-\infty}{#1}}
\newcommand\tot{\sif{t}}
\newcommand\Vko[2]{\cV_k({#1},{#2},\omega)}
\newcommand\Vkosyn[2]{\cV_k^{(syn)}({#1},{#2},\omega)}
\newcommand\Vkoext[2]{\cV_k^{(ext)}({#1},{#2},\omega)}
\newcommand\Vkodet[2]{\cV_k^{(det)}({#1},{#2},\omega)}
\newcommand\Vkdet[1]{\cV_k^{(det)}({#1})}
\newcommand\Vldet[1]{\cV_l^{(det)}({#1})}
\newcommand\Vkonoise[2]{\cV_k^{(noise)}({#1},{#2},\omega)}
\newcommand\VkoB[2]{\cV_k^{(B)}({#1},{#2},\omega)}
\newcommand\gk[1]{g_k\pare{#1}}
\newcommand\Vk[1]{\cV_k({#1})}
\newcommand\Xk[1]{X_k({#1})}
\newcommand\dXk[1]{\delta X_k\pare{#1}}
\newcommand\Repk[2]{\cR^{({#1})}_k\pare{#2}}
\newcommand\zXk[1]{X^{(sp)}_k\pare{#1}}
\newcommand\Vkop[1]{\cV_k({#1},\omega')}
\newcommand\Xkop[1]{X_k({#1},\omega')}
\newcommand\sko[2]{\sigma_k({#1},{#2},\omega)}
\newcommand\sr{\sigma_R}
\newcommand\sdr{\sigma^2_R}
\newcommand\sk[1]{\sigma_k({#1})}
\newcommand\skd[1]{\sigma^2_k({#1})}
\newcommand\skop[1]{\sigma_k({#1},\omega')}
\newcommand\skopd[1]{\sigma^2_k({#1},\omega')}
\newcommand\tauk[1]{\tau_k({#1})}
\newcommand\tko{\tau_k(t,\omega)}
\newcommand\tlo{\tau_l(t,\omega)}
\newcommand\nko{\tau_k(n,\omega)}
\newcommand\tkop{\tau_k(t,\omega')}
\newcommand\nkop{\tau_k(n,\omega')}
\newcommand\Vksyn[1]{\cV_k^{(syn)}({#1})}
\newcommand\Vksynp[1]{\cV_k^{(syn)}({#1},\omega')}
\newcommand\Vkext[1]{\cV_k^{(ext)}({#1})}
\newcommand\Vkextp[1]{\cV_k^{(ext)}({#1},\omega')}
\newcommand\Vknoise[1]{\cV_k^{(noise)}({#1})}
\newcommand\VkB[1]{\cV_k^{(B)}({#1})}
\newcommand\Vkto{\Vko{t}{t+1}}
\newcommand\Vksto{\Vko{s}{t}}
\newcommand\tLk{\tau_{L,k}}
\newcommand\tmk{\tau_{M,k}}
\newcommand\gmk{g_{M,k}}
\newcommand\gLk{g_{L,k}}
\newcommand\vm[1]{var_m\left[#1 \right]}
\newcommand\dl[1]{\delta_l\left[#1 \right]}
\newcommand\eg[1]{\stackrel {\rm #1}{=}}
\newcommand\nega[1]{\stackrel {\rm #1}{\neq}}
\newcommand\akj[1]{\alpha_{kj}\pare{#1}}
\newcommand\moy[1]{\mu\bra{#1}}
\newcommand\moyc[2]{\mu\bra{#1 \, | \, #2}}
\newcommand\tmoy[1]{\tmu\bra{#1}}
\newcommand\tmoyc[2]{\tmu\bra{#1 \, | \, #2}}
\newcommand\M[1]{\cM^{(#1)}}
\newcommand\Mc[3]{\cM^{(#1)}_{#2,\,#3}}
\newcommand\p[1]{p^{(#1)}}
\newcommand\pc[2]{p^{(#1)}_{#2}}
\newcommand\tphi[1]{\tilde{\phi}^{(#1)}}
\newcommand\tfi{\tilde{\phi}}
\newcommand\tphic[3]{\tilde{\phi}^{(#1)}_{#2,\,#3}}
\newcommand\tf[1]{\tilde{f}^{(#1)}}
\newcommand\tfc[2]{\tilde{f}^{(#1)}_{#2}}
\newcommand\tmu{\tilde{\mu}}
\newcommand\cst[3]{#1_{#2}^{(#3)}}
\newcommand\spont[2]{#1_{(sp)}\bra{#2}}
\newcommand\spontc[3]{#1_{(sp)}\brac{#2}{#3}}

\begin{document}
\begin{bmcformat}


\title{Statistics of spike trains in conductance-based neural networks: Rigorous results.}
 

\author{B. Cessac\correspondingauthor$^{1}$%
       \email{bruno.cessac@inria.fr}%
      }


\address{%
    \iid(1)NeuroMathComp, INRIA, 2004 Route des Lucioles, 06902 Sophia-Antipolis, France
}%

\maketitle


\begin{abstract}
We consider a conductance based neural network inspired by the generalized Integrate 
and Fire model introduced by Rudolph and Destexhe in \cite{rudolph-destexhe:06}.
We show the existence and uniqueness of a unique Gibbs distribution characterizing
spike train statistics. The corresponding Gibbs potential is explicitly computed.
These results hold in presence of a time-dependent stimulus and apply therefore to non-stationary dynamics. 
 \end{abstract}

\ifthenelse{\boolean{publ}}{\begin{multicols}{2}}{}


\su{Introduction}
Neural networks have an overwhelming complexity. While an isolated
neuron can exhibit a wide variety of responses to stimuli \cite{izhikevich:04},
from regular spiking to chaos \cite{guckenheimer-oliva:02,faure-korn:97},
neurons coupled in a network via synapses (electrical or chemical)
may show an even wider variety of collective dynamics \cite{faure-korn:01}
resulting from the conjunction of nonlinear effects, time propagation delays,
synaptic noise, synaptic plasticity, and external stimuli \cite{ermentrout-terman:10}. Focusing on the action potentials, this complexity is manifested by drastic
changes in the spikes activity, for instance when switching from spontaneous to evoked
activity (see for example A. Riehle's team experiments on the monkey motor cortex
\cite{riehle-grun-etal:97,grammont-riehle:99,riehle-etal:00,grammont-riehle:03}).
However, beyond this complexity may exist some hidden laws ruling an (hypothetical)
``neural code'' \cite{rieke-warland-etal:97}. 

One way of unraveling these hidden laws is to seek some regularities or reproducibility in the statistics of spikes.
 While early investigations on spiking activities were focusing on firing rates where neurons are considered as independent sources, researchers concentrated more recently on collective statistical indicators such as pairwise correlations. Thorough experiments  in the retina
\cite{schneidman-berry-etal:06,tkacik-schneidman-etal:09} as well as in the parietal cat cortex 
\cite{marre-boustani-etal:09}, suggested that such correlations are  crucial for understanding spiking activity.
Those conclusions where obtained
using the \textit{maximal entropy principle} \cite{jaynes:57}. Assume that
the average value of observables quantities (e.g. firing rate or spike correlations)
has been measured. Those average values constitute constraints for the statistical model.
In the maximal entropy principle, \textit{assuming stationarity}, one looks for the probability distribution which maximizes the statistical entropy given those constraints. This leads to a (time-translation invariant) Gibbs distribution.
In particular, fixing firing rates and the probability of pairwise coincidences of spikes leads
to a Gibbs distribution having the same form as the Ising model.
This idea has been introduced by Schneidman et al in 
\cite{schneidman-berry-etal:06} for the analysis of retina spike trains.
They reproduce accurately the probability of
\textit{spatial} spiking pattern. Since then, their approach has known a great success
(see e.g. \cite{shlens-field-etal:06,cocco-leibler-etal:09,tang-jackson-etal:08}), although some authors raised
solid objections on this model \cite{roudi-nirenberg-etal:09,shlens-field-etal:09,tkacik-schneidman-etal:09,ohiorhenuan-mechler-etal:10} while several papers have pointed out the importance of \textit{temporal} patterns of activity at the network level \cite{lindsey-morris-etal:97,villa-tetko-etal:99,segev-baruchi-etal:04}. 
As a consequence, a few authors \cite{marre-boustani-etal:09,amari:10,roudi-hertz:10}
have attempted to define time-dependent models of Gibbs distributions where constraints include time-dependent correlations
between pairs, triplets and so on \cite{vasquez-vieville-etal:11}. As a matter of fact, the analysis
of the data of \cite{schneidman-berry-etal:06} with such models describes more accurately the statistics
of \textit{spatio-temporal} spike patterns  \cite{vasquez-palacios-etal:11}. \\

Taking into account all constraints inherent to experiments it seems extremely difficult to find an optimal model describing spike trains statistics. It is in fact likely that there is not one model, but many, depending on the experiment, the stimulus, the investigated part of the nervous system and so on. Additionally, the assumptions made in the works
quoted above are difficult to control. Especially, the maximal entropy principle assumes a stationary dynamics while many
experiments consider a time-dependent stimulus generating a time-dependent response
where the stationary approximation may not be valid. At this stage, having an example
where one knows the explicit form of the spike trains probability distribution  would be helpful to control those assumptions
and to define related experiments. 

This can be done considering neural network models. Although, to be tractable, such models may be quite away from biological plausibility, they can give  hints on which statistics can be expected in real neural networks. But, even in the simplest examples, characterizing spike statistics arising from  the conjunction of nonlinear effects, time propagation delays,
synaptic noise, synaptic plasticity, and external stimuli is far from being trivial
on mathematical grounds.

In \cite{cessac:10b} we have nevertheless proposed an exact and explicit result for the characterization of spike trains statistics in a discrete time version of Leaky Integrate-and-Fire neural network.  The results were quite surprising. It has been shown that
whatever the parameters value (in particular synaptic weights), spike trains are distributed according to a Gibbs distribution whose potential can be explicitly computed.
The first surprise lies in the fact 
that this potential has infinite range,
namely spike statistics has an infinite memory.
This is because the membrane potential evolution integrates its past values and the past influence of the network via the leak term.
Although Leaky Integrate-and-Fire models have a reset mechanism which erases the memory of the neuron
whenever it spikes, it is not possible to upper bound the next time of firing. As a consequence, statistics is non-Markovian
(for recent examples of non-Markovian behavior in neural models see also \cite{kravchuk-vidybida:10}). 
The infinite range of the potential corresponds, in the maximal entropy principle interpretation, to having infinitely many constraints.\\

Nevertheless, the leak term  influence decays exponentially fast with time (this property guarantees the existence and uniqueness of a Gibbs distribution).  
As a consequence, one can approximate the exact Gibbs distribution by the invariant probability of a Markov chain, with a memory depth proportional to the log of the (discrete time) leak term. In this way, the truncated potential corresponds to a finite number of constraints
in the maximal entropy principle interpretation. However,
the second surprise is that  this approximated potential is nevertheless  far from the Ising model or any of the models discussed above, that appear as quite bad approximations. In particular, there is a need
 to consider
$n$-uplets of spikes with time delays. This mere fact asks hard problems about evidencing such type of potentials in experiments. Especially, new type of algorithms for spike trains analysis have to be developed \cite{vasquez-vieville-etal:10}. \\   

The model considered in \cite{cessac:10b} is rather academic: time evolution is discrete, synaptic interactions are instantaneous, dynamics is stationary (the stimulus is time-constant) and, as in a Leaky Integrate-and-Fire model, conductances are constant. It is therefore necessary to investigate whether our conclusions remain for more realistic
neural networks models. In the present paper we consider a conductance-based model
introduced by Rudolph and Destexhe in \cite{rudolph-destexhe:06} called ``generalized Integrate and Fire'' (gIF) model. This model allows one to consider 
realistic synaptic responses, and conductances depending on spikes arising
in the past of the network, leading to a rather complex dynamics
which has been characterized in \cite{cessac-vieville:08} in the deterministic
case (no noise in the dynamics). Moreover, the biological plausibility of this model
is well accepted \cite{jolivet-lewis-etal:04,jolivet-rauch-etal:06}.

Here we analyse spike statistics in the gIF model with noise and with a time-dependent stimulus. Moreover, the post-synaptic potential profiles are quite general and summarize all the examples that we know in the literature. Our main result is to prove the existence and uniqueness of a Gibbs measure characterizing spike trains statistics, for all parameters compatible with physical constraints (finite synaptic weights, bounded stimulus,
and positive conductances). Here, as in \cite{cessac:10b}, the corresponding Gibbs potential has infinite range corresponding to a non-Markovian dynamics, although Markovian approximations can be proposed in the gIF model too. The Gibbs potential depends on all parameters in the model (especially connectivity and stimulus) and has a form
quite more complex than Ising-like models. As a by-product of the proof of our main result, additional interesting notions and results are produced such as    
continuity with respect to a raster, or  exponential decay of memory thanks to the shape of synaptic responses.\\

The paper is  organised as follows. In the  section \ref{SIF} we briefly introduce Integrate-and-Fire models and propose two important extensions of the classical models: the spike has a duration and the membrane potential is reset to a non-constant value. These extensions, which are necessary for the validity of our mathematical results,
 render nevertheless the model more biologically plausible (see the discussion section). One of the keys of the present work is to consider spike trains (raster plots) as infinite sequences. Since in gIF models conductances are updated upon the occurrence of spikes,
one has to consider two types of variables with distinct type of dynamics.
On one hand, the membrane potential, which is the physical variable associated with
neurons dynamics evolves continuously. 
On the other hand, spikes are discrete events. 
Conductances are updated according to these discrete time events. 
The formalism introduced in section \ref{SIF} and \ref{SgIF} 
allows us to handle properly this mixed dynamics. As a consequence, these sections define gIF model with more mathematical structure than the original paper \cite{rudolph-destexhe:06} and mostly contain original results. 
Moreover, we add to the model several original features such as the consideration of a general form of synaptic profile with exponential decay or the introduction of noise. Section \ref{SgIF_fixed_rast} proposes a preliminary analysis of gIF model-dynamics. In sections \ref{Sbounds}, \ref{Scont} we provide several useful mathematical propositions as a necessary step toward the analysis of spike statistics, developed in section \ref{Sstat}, where we prove the main result of the paper: existence and uniqueness of a Gibbs distribution describing spike statistics.  
The section \ref{Scons} and \ref{SDisc} are devoted to a  discussion on practical consequences of our results for neuroscience.

\su{Integrate and Fire model.}\label{SIF}

We consider the  evolution of a set of $N$ neurons.
Here, neurons are considered as ``points'' instead of
spatially extended and structured objects. As a consequence, 
we define, for each neuron $k \in \left\{1   \dots N \right\}$,
 a variable $V_k(t)$ called the ``membrane potential of neuron $k $ at time
$t$'' without specification of which part of a real neuron 
(axon, soma, dendritic spine, ...) it corresponds to. 
Denote $V(t)$  the vector $\pare{V_k(t)}_{k=1}^N$. 

We focus here on ``Integrate-and-Fire models'',
where dynamics always consists of two regimes.

\ssu{ The ``Integrate regime''.} 
Fix a real number $\theta$ called the ``firing threshold of the neuron''\footnote{We  assume that all neurons have the
same firing threshold.
The notion of threshold is already an approximation which 
is not sharply defined in Hodgkin-Huxley \cite{hodgkin-huxley:52} or Fitzhugh-Nagumo \cite{fitzHugh:61,nagumo-etal:62} models (more precisely
it is not a constant but it depends on the dynamical variables \cite{cronin:87}). 
Recent experiments \cite{naundorf-etal:06,cormick-etal:07,naundorf-etal:07} even suggest
that there may be no real potential threshold. \label{Fthresh}}.
Below the threshold, $V_k<\theta$, neuron $k$'s dynamics 
is driven by an equation of the form: 

\beq \label{Integrate}
C_k\frac{dV_k}{dt}+g_k V_k=i_k,
\eeq

\nid where $C_k$ is the membrane capacity of neuron $k$. 
In its most general form, the neuron $k$'s membrane conductance $g_k > 0$ depends on $V_k$ plus additional variables
such as the probability of having ionic channels open (see e.g. Hodgkin-Huxley equations \cite{hodgkin-huxley:52})
as well as on time $t$. The explicit form of $g_k$ in the present model is developed in section \ref{Ssyn}.
The current $i_k$ typically depend  on time $t$, and on the past activity of the network. 
It also contains a stochastic component modelling noise in the system (e.g. synaptic transmission, see section \ref{Snoise}).

\ssu{LIF model}
A classical example of Integrate-and-Fire model is the Leaky Integrate-and-Fire's (LIF) introduced in \cite{lapicque:07}
where equation (\ref{Integrate}) reads:

\beq\label{LIF}
\frac{dV_k}{dt}=-\frac{V_k}{\tau_L}+\frac{i_k(t)}{C_k}.
\eeq

\nid where $g_k$ is a constant
 and $\tau_L=\frac{C_k}{g_k}$
is the characteristic time for membrane potential
decay when no current is present (``leak term'').

\ssu{Spikes}
The dynamical evolution (\ref{Integrate}) may eventually lead $V_k$
to exceed $\theta$. If, at some time $t$, $V_k(t) = \theta$ then neuron $k$ emits
a spike or ``fires''.  In our model, like in biophysics, a spike has a finite duration $\delta > 0$;
this is a generalisation of the classical formulation of Integrate-and-Fire models where the spike
 is considered instantaneous.  On biophysical grounds $\delta $ 
is of order of a  millisecond. Changing the time units we 
may set $\delta=1$ without loss of generality. 
Additionally, neurons have a refractory period 
$\tau_{refr}>0$ where they are not able 
to emit a new spike although their membrane potential
can fluctuate below the threshold (see fig. \ref{Fspike}). 
Hence, spikes emitted by a given neuron are separated
by a minimal time scale
\beq\label{tausep}
\tau_{sep}=\delta+\tau_{refr}
\eeq

\ssu{Raster plots}

In experiments spiking neurons activity is represented by ``raster plots'',
namely a graph with time in abscissa and a neuron labeling in ordinate such
that a vertical bar is drawn each ``time'' a neuron emits a spike.
Since spikes have a finite duration $\delta$
such a representation limits the time resolution: events
with a time scale smaller that $\delta$ are not distinguished.
As a consequence, if neuron $1$ fires at time $t_1$ and
neuron $2$ at time $t_2$ with $\abs{t_2-t_1} < \delta=1$ the two
spikes appear to be simultaneous on the raster.
Thus, the raster representation introduces a time quantization
and has a tendency to enhance synchronization.
In gIF models conductances are updated upon the occurrence of spikes 
(see section \ref{Scond_unitary}) which are considered as such discrete events.
This could correspond to the following ``experiment''. Assume that we measure
the spikes emitted by a set of in vitro neurons, and that we use this information
to update the conductances of a  model, in order to see how this model
``matches'' the real neurons (see \cite{jolivet-etal:06} for a nice investigation in this
spirit). Then, we would have to take into account that the information provided
by the experimental raster plot is discrete, even if the
membrane potential evolves continuously. 
The consequences of this time-discretisation as well as the limit $\delta \to 0$ are
developed in the discussion section.

As a consequence, one has to consider two types of variables with distinct type of dynamics.
On one hand, the membrane potential, which is the physical variable associated with
neuron dynamics evolves with a continuous time. 
On the other hand, spikes, which are the quantities of interest in the present paper are discrete events. 
To  define properly this mixed dynamics and study its properties
we have to model  spikes times
and raster plots.

\ssu{Spike times}
If, at time $t$, $V_k(t)=\theta$, a spike is registered
at the \textit{integer time immediately after $t$}, called the spike time. Choosing integers
for the spike time occurrence is a direct consequence of setting $\delta=1$. 
Thus,
to each neuron $k$ and integer $n$
we associate a ``spiking state'' defined by: 
$$
\omega_k(n) = 
\left\{
\baR{ccc}
1 & \quad \mbox{if} \quad \exists t \in ]n-1,n] \quad \mbox{such \, that} \quad V_k(t) = \theta;\\
&&\\
0 & \quad \mbox{otherwise.}
\eaR
\right.
$$
For convenience
and in order to simplify the notations in the mathematical 
developments, we call $\ent{t}$  the largest integer which is $ \leq t$
(thus $\ent{-1.2}=-2$ and $\ent{1.2}=1$). Thus, the integer immediately after $t$ is $\ent{t+1}$ and 
we have therefore 
that $\omega_k(\ent{t+1})=1$ whenever $V_k(t) = \theta$.
Although, characteristic events in a raster plot are spikes (neuron fires)
it is useful in subsequent developments to consider also
the case when neuron is not firing ($\omega_k(n)=0$). \\

\ssu{Reset}
In the classical formulation of Integrate-and-Fire models the spike
occurs \textit{simultaneously} with a reset of the membrane potential
to some \textit{constant} value $\Vr$, called the ``reset potential''.
Instantaneous reset is a source of pathologies as discussed in \cite{cessac-vieville:08,cessac:10a} and in the discussion section.
Here, we consider that reset occurs after the time delay 
$\tau_{sep} \geq 1$ including spike duration and refractory period. 
We set:
\beq\label{Fire}
V_k(t) = \theta \Rightarrow V_k(\ent{t+\tau_{sep}}) = \Vr.
\eeq
The reason why the reset time is the integer number $\ent{t+\tau_{sep}}$
instead of the real $t+\tau_{sep}$ is that it eases the notations and proofs. 
Since the reset value is random (see below and Fig. \ref{Fspike}) this assumption
has no impact on the dynamics. 

Indeed, in our model, the reset value $\Vr$ is not a constant. This is
  a Gaussian random variable with mean zero (we set the rest potential
to zero without loss of generality) and variance
$\sdr >0$. In this way we model the spike duration and refractory period,
as well as the random oscillations of the membrane potential during the refractory
period.  As a consequence, the value
of $V_k$ when the neuron can fire again is not a constant, as it is
in classical IF models. 
A related reference (spiking neurons with partial reset) 
is \cite{kirst-geisel-etal:09}. The assumption that $\sdr >0$ is necessary for our mathematical
developments (see the bounds (\ref{bounds_sigma_k})). 
We assume $\sdr$ to be small to avoid trivial and unrealistic situations
where $\Vr \geq \theta$ with a large probability leading the neuron to fire all the time. Note however that this is not a required assumption to establish our mathematical results. We also assume that, in successive resets, the random variables
$\Vr$ are independent.

\ssu{The shape of membrane potential during the spike}

On biophysical grounds the time course of the membrane potential during the spike includes a depolarisation and re-polarisation phase 
due to the nonlinear effects of gated ionic channels on the conductance. This leads to introduce, in modelling, additional variables such as activation/inactivation probabilities as in the Hodgkin-Huxley model \cite{hodgkin-huxley:52} or adaptation current as e.g. in FitzHugh-Nagumo model \cite{fitzHugh:55,fitzHugh:61,nagumo-etal:62,fitzhugh:69} (see the discussion section
for extensions of our results to those models). Here, since we are considering only one variable for the neuron state, the membrane potential,
  we need to define the spike profile, i.e. the  course
of $V_k(t)$ during the time interval $(t,\bra{t+\tau_{sep}})$. It turns out that the precise
shape of this profile plays no role in the developments proposed here, where
we concentrate on spike statistics. 
Indeed, a spike is registered whenever $V_k(t)=\theta$ and this does not depend
on the spike shape. What we need is therefore to define the  membrane potential evolution
before the spike, given by  (\ref{Integrate}), and after the spike, given by  (\ref{Fire})
(see Figure \ref{Fspike}). 

\begin{figure}[h!]
\begin{center} 
\includegraphics[height=6cm,width=8cm]{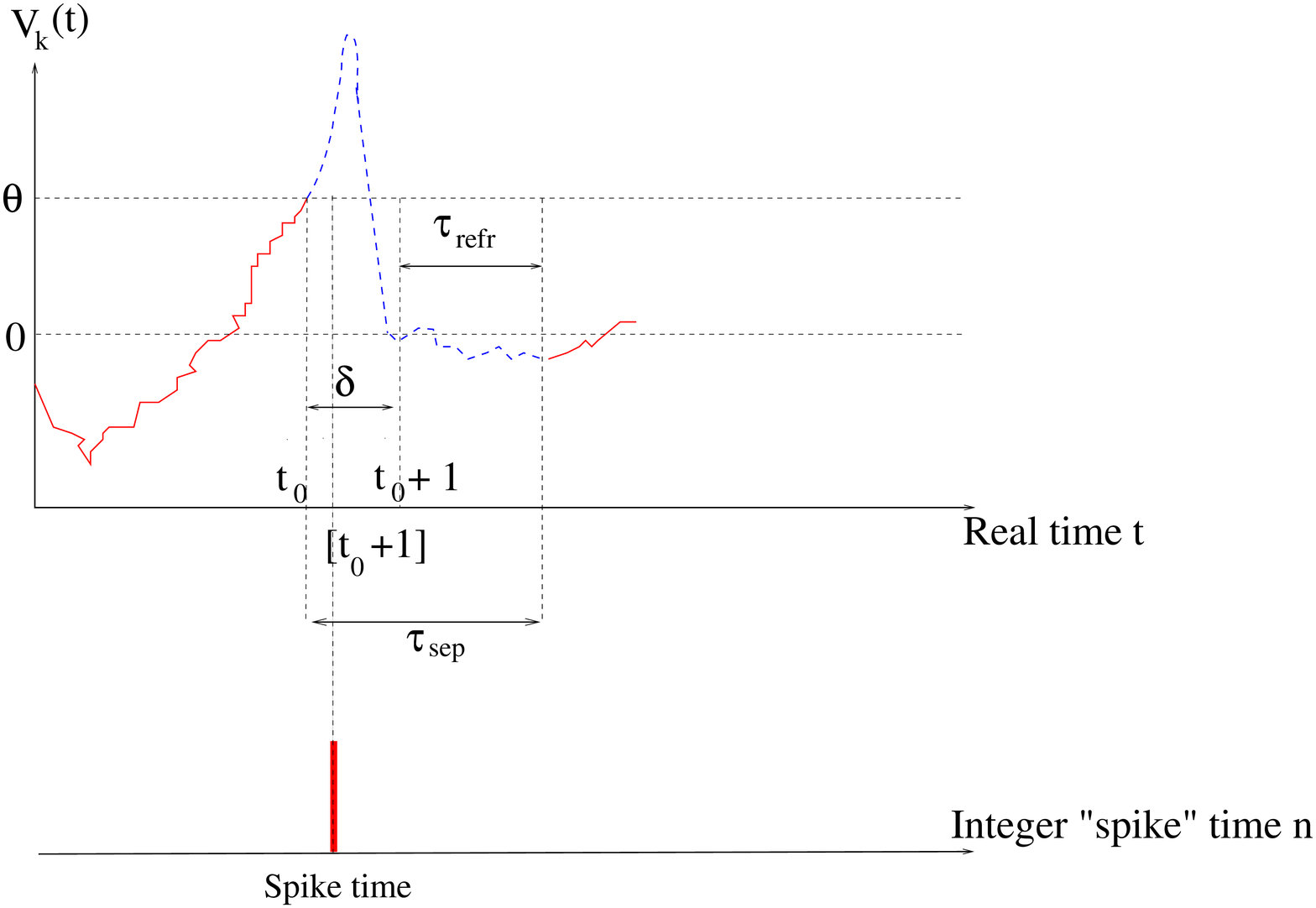} 
\end{center}
\caption{Time course of the membrane potential in our model. The blue dashed curve illustrates the shape
of a real spike, but what we model is the red curve.}
\label{Fspike}
\end{figure}

\ssu{Mathematical representation of raster plots}\label{Sraster}

The ``spiking pattern'' of the neural network at integer time $n$  is the vector 
$\omega(n)=\left(\omega_k(n)\right)_{k=1}^N$. For $m < n$ we note  
$\seq{\omega}{m}{n} = \Set{\omega(m) \, \omega(m+1) \dots \omega(n)}$ 
the ordered sequence of spiking patterns  between $m$ and $n$. Such sequences are
 called \textit{spike blocks}.
Additionally we note $\seq{\omega}{m}{n_1-1}\seq{\omega}{n_1}{n}=\seq{\omega}{m}{n}$ the concatenation
of the blocks $\seq{\omega}{m}{n_1-1}$ and $\seq{\omega}{n_1}{n}$.

Call $\cA = \Set{0,1}^N$ the set of spiking patterns 
(alphabet). An element of  $X \deq \cA^\setZ$,
i.e. a bi-infinite ordered sequence $\tom=\left\{\omega(n)\right\}_{n=-\infty}^{+\infty}$ 
of spiking patterns, is called a 
 ``raster plot''. It tells us which neurons are firing at each 
time $n \in \setZ$. In  experiments raster plots are
obviously finite sequences of spiking pattern but the extension
to $\setZ$, especially the possibility of considering an arbitrary
distant past (negative times) is a key of the present
work. In particular, the notation $\sif{n}$ refers to spikes occurring
from $-\infty$ to $n$.\\

To each raster $\omega \in X$ and each neuron index\footnote{We use the following convention. The index $k$ is used for a post-synaptic neuron while the index $j$ refers to pre-synaptic neurons. Spiking events are used to update the conductance of neuron $k$ according to spikes emitted by pre-synaptic neurons. That's why we label the spike times with an index $j$.} $j =1 \dots N$ we associate
an ordered  (generically infinite) list of ``spike times'' $\Set{\tjro}_{r=1}^\infty$ (integer numbers) such that $\tjro$ is 
the $r$-th time of firing of neuron $j$ in the raster $\omega$.
In other words, we have $\omega_j(n)=1$ if and only if 
$n = \tjro$ for some $r=1, \dots +\infty$. 

We introduce here two specific rasters which are of use in the paper. We note
$\Omega_0$ the raster  such that $\omega_k(n)=0, \, \forall k=1 \dots N, \forall n \in \setZ$
(no neuron ever fires) and $\Omega_1$ the raster $\omega_k(n)=1, \, \forall i=1 \dots N, \forall n \in \setZ$
(each neuron is firing at every integer time).

Finally, we use the following notation borrowed from \cite{maillard:07}.
We note, for $n \in \setZ$, $m \geq 0$, and $r$ integer:
\beq\label{eg}
\omega \eg{m,n} \omega' \quad \mbox{if} \quad\omega(r)=\omega'(r), \, \forall r \in \Set{n-m, \dots, n}.
\eeq

For simplicity, we consider that $\tau_{ref}$, the refractory period,
is smaller than $1$ so that a neuron can fire two consecutive time steps
(i.e. one can have $\omega_k(n)=1$ and $\omega_k(n+1)=1$). This constraint is
discussed in section \ref{SGrammar}. 

\ssu{Representation of time dependent functions}

Throughout the paper we use the following convention.
For a real function of $t$ and $\omega$, we write
$f(t,\omega)$ for $f(t,\sif{\ent{t}})$ to simplify notations.  This notation takes
into account the duality between variables such as membrane potential
evolving with respect to a continuous time and raster plots labeled with discrete time.
Thus, the function $f(t,\omega)$ is a function of the continuous variable $t$ and of
the spike block $\sif{\ent{t}}$, where by definition $\ent{t} \leq t$, namely $f(t,\omega)$
depends on the spike sequences occurring \textit{before} $t$. This constraint is imposed
by causality.

\ssu{Last reset time}\label{LastReset}

We define  $\tko$ as the
 last time before $t$ where neuron $k$'s membrane potential  has been reset, in the raster $\omega$. This is $-\infty$
if the membrane potential has never been reset. 
As a consequence of our choice (\ref{Fire}) for the reset time
 $\tko$ is an integer number fixed by $t$ and the raster before $t$ . 
The membrane potential value of neuron $k$ at time $t$
is controlled by the reset value $\Vr$ at time $\tko$ and by
the further sub-threshold evolution (\ref{Integrate}) 
from time $\tko$ to time $t$.

\su{Generalized Integrate-and-Fire models}\label{SgIF}

In this paper, we concentrate on an extension 
of (\ref{LIF}), called ``generalized Integrate-and-Fire'' (gIF), introduced in \cite{rudolph-destexhe:06}, closer to
 biology \cite{jolivet-lewis-etal:04,jolivet-rauch-etal:06}, since it considers
more seriously neurons interactions via synaptic responses.

\ssu{Synaptic conductances}\label{Ssyncond}

Depending on the neuro-transmitter\footnote{AMPA, NMDA, GABA A,  GABA B \cite{destexhe-mainen-etal:98}.}
 they use for synaptic transmission
neurons can be excitatory (population $\cE$) or inhibitory (population $\cI$).
This is modeled by introducing reversal potentials  
$E^+$ for excitatory (typically $E^+ \simeq 0 mV$ for AMPA and NMDA) and $E^-$ for inhibitory ($E^- \simeq -70 mV$ for GABA A and
$E^- \simeq -95 mV$ for GABA B). We focus here on one population of excitatory and one population of inhibitory neurons although
extensions to several populations may be considered as well.
Also, each neuron is submitted to a current $I_k(t)$.
We assume that this current has some stochastic component that mimics synaptic noise (section \ref{Snoise}).

The variation of the membrane potential of neuron $k$ at time $t$
reads:

\beq\label{diff_gen}
C_k \, \frac{dV_k}{dt}=
-\gLk \, (V_k -E_L) 
\, - g^{(\cE)}_k(t) \, (V_k-E^+)  
\, - g^{(\cI)}_k(t) \, (V_k-E^-)
\, + I_k(t),
\eeq
where $\gLk$ is a leak conductance, $E_L < 0$ is the leak reversal potential (about $-65$ mV),
 $g^{(\cE)}_k(t)$
the conductance of the excitatory population
and $g^{(\cI)}_k(t)$ the conductance of inhibitory population.
They are given by:

\beq\label{cond_syn}
g^{(\cE)}_k(t) = \sum_{j \in \cE} g_{kj}(t); \quad
g^{(\cI)}_k(t) = \sum_{j \in \cI} g_{kj}(t),
\eeq
where $g_{kj}$ is the conductance of the synaptic contact $j \to k$.

We may rewrite  equation (\ref{diff_gen}) in the form (\ref{Integrate}) setting
$$
g_k(t)=\gLk \, + \, g^{(\cE)}_k(t) \, + \, g^{(\cI)}_k(t),
$$
and
$$
i_k(t)=\gLk \, E_L 
+  g^{(\cE)}_k(t) \, E^+
+ g^{(\cI)}_k(t) \, E^-
\, + \, I_k(t).
$$

\ssu{Conductance update upon a spike occurrence.}\label{Scond_unitary}

The conductances $g_{kj}(t)$ in (\ref{cond_syn}) depend  on time $t$ but also on 
pre-synaptic spikes occurring before $t$. This is a general statement
which is modeled in gIF models as follows. 
Upon arrival of a spike in the pre-synaptic neuron $j$ at time $\tjro$
the membrane conductance of the post-synaptic neuron $k$ is modified
as:
\beq\label{cond_update}
g_{kj}(t) \, = \, g_{kj}(\tjro) \, + \, G_{kj} \, \akj{t \, - \, \tjro}, \quad t > \tjro.
\eeq

In this equation,  the quantity $G_{kj} \geq 0$ characterizes the maximal 
amplitude of the conductance during a post-synaptic potential. We use the convention that $G_{kj} \, = \,0$
if and only if there is no synapse between $j$ and $k$. This allows us to encode
the graph structure of the neural network in the matrix $G$ with entries $G_{kj}$.
Note that the $G_{kj}$'s can evolve in time due to synaptic plasticity mechanisms
(see section \ref{Ssynplast}).\\

The function $ \alpha_{kj}$ (called ``alpha'' profile \cite{destexhe-mainen-etal:98}) mimics 
the time course of the synaptic conductance upon the occurrence of the spike. 
Classical examples are:  

\beq\label{alpha_exp}
\akj{t}=  e^{-\frac{t}{\tau_{kj}}}  \, H(t),
\eeq
(exponential profile)
or:
\beq\label{alpha_t_exp}
\akj{t}=  \frac{t}{\tau_{kj}} \, e^{-\frac{t}{\tau_{kj}}} \, H(t),
\eeq
\nid with $H$ the Heaviside function (that mimics causality) 
and $\tau_{kj}$ is the   characteristic decay times of the synaptic response. Since $t$ is a time, the division by $\tau_{kj}$ ensures that $\akj{t}$ is a dimensionless quantity: this eases the legibility of the subsequent equations on physical grounds (dimensionality of physical quantities).
  
Contrarily to (\ref{alpha_exp}) the synaptic profile (\ref{alpha_t_exp}),
 with $\akj{0}=0$ while $\akj{t}$ is maximal
for $t=\tau_{kj}$ allows one to smoothly delay the spike action on the post-synaptic
neuron. 
More general forms of synaptic responses
 could be considered as well \footnote{For example,
 the $\alpha$ profile may obey a Green equation of type \cite{ermentrout:98}:
$$
\sum_{l=0}^k a_{kj}^{(l)}  \frac{d^l \alpha_{kj}}{d u^l}(t)=\delta(t),
$$
where $k=1$, $a_{kj}^{(0)}=\frac{1}{\tau_{kj}}$, $a_{kj}^{(1)}=1$, corresponds to (\ref{alpha_exp}), and so on.}.

\ssu{Mathematical constraints on the synaptic responses}

In all the paper we assume that the $\alpha_{kj}$'s are positive and bounded. Moreover, we assume that:
\beq\label{alpha_as}
\akj{t} \sim \frac{t^d}{\tau_{kj}^d} e^{-\frac{t}{\tau_{kj}}} \deq f_{kj}(t), \quad t \to +\infty,
\eeq
for some integer $d$. So that $\akj{t}$ decays exponentially fast as $t \to +\infty$, with a characteristic time
 $\tau_{kj}$, the decay time of the evoked post-synaptic potential. 
This constraint matches all synaptic response kernels  that we know (where typically $d=0,1$)
\cite{destexhe-mainen-etal:98,ermentrout:98}.
 
This has the following consequence. For all $t$, 
$M<t$ integer, $r$ integer, we have, setting $t=\frc{t}+\ent{t}$,
where $\frc{t}$ is the fractional part:
$$
\sum_{r < M} \akj{t-r} =
\sum_{r <M} \akj{\ent{t}-r+\frc{t}}= 
\sum_{n> \ent{t}-M} \akj{n+\frc{t}}.
$$
 Therefore, as $M \to -\infty$,
$$
\sum_{r < M} \akj{t-r} \sim 
\sum_{n \geq \ent{t}-M+1} f_{kj}(n+\frc{t})
<\int_{t-M}^{+\infty} f_{kj}(u) \, du
=
P_d\pare{\frac{t-M}{\tau_{kj}}} \, e^{-\frac{t-M}{\tau_{kj}}},
$$
where $P_d()$ is a polynomial of degree $d$.

We introduce the following (Hardy) notation:
if a function $f(t)$ is bounded from above, as $t \to +\infty$,
by a function $g(t)$  we write:
$f(t) \preceq g(t)$. Using this notation we have therefore:
\bp\label{Llim_alpha}
\beq\label{lim_alpha}
\sum_{r < M} \akj{t-r} \preceq
P_d\pare{\frac{t-M}{\tau_{kj}}} \, e^{-\frac{t-M}{\tau_{kj}}},
\eeq
\ep
as $M \to -\infty$.\\
  
Additionally, the constraint (\ref{alpha_as}) 
implies that there is some $\alpha^+ < +\infty$
such that, for all $t$, for all $k,j$:
\beq\label{int_alpha}
\sum_{r < t} \akj{t-r} \leq \alpha^+.
\eeq
Indeed, for $n \geq 0$ integer, call $A_{kj}(n)=\sup\Setc{ \alpha_{kj}(n+x)}{x \in [0,1[}$. Then, 
$$\sum_{r < t} \akj{t-r} 
= \sum_{n=1}^{+\infty} \akj{n+\frc{t}}
 \leq \sum_{n=1}^{+\infty}A_{kj}(n).$$
Due to (\ref{alpha_as}) this series converges (e.g. from Cauchy criterion). We set:
$$\alpha_+=\max_{kj} \sum_{n=1}^{+\infty} A_{kj}(n).$$

On physical grounds it implies that the conductance $g_k$
remains bounded, even if each pre-synaptic neuron is firing all the time
(see eq. (\ref{bounds_gk}) below). \\

\ssu{Synaptic summation}\label{Ssyn}

Assume that eq. (\ref{cond_update}) remains valid for an arbitrary number of pre-synaptic spikes emitted
by neuron $j$ within a finite time interval $[s,t]$ (i.e. neglecting nonlinear effects such as the fact that there
is a finite amount of neurotransmitter leading to saturation effects).
Then, one obtains the following equation
for the conductance $g_{kj}$ at time $t$, upon the arrival of spikes
at times $\tjro$ in the time interval $[s,t]$:  
$$g_{kj}(t)=  g_{kj}(s) +  G_{kj} \, 
\sum_{ \Setc{r}{s \leq \tjro < t}} \akj{t \, - \,  \tjro}.$$

The conductance at time $s$, $g_{kj}(s)$, depends on the neuron $j$'s activity preceding $s$.
This term is therefore unknown unless one knows exactly the past evolution before $s$.
One way to circumvent this problem is to
taking $s$ arbitrary far in the past, i.e. taking $s \to - \infty$  in order to remove the dependence on initial conditions. This  corresponds to the following situation. When one observes a real neural network the time where the observation starts, say $t=0$, is usually not the time when the system has begun to exist, $s$ in our notations. Taking $s$ arbitrary far in the past corresponds to assuming that the system has evolved long enough so that it has reached sort of a ``permanent regime'', not necessarily stationary, when the observation starts. On phenomenological grounds it is enough to take $-s$ larger than all characteristic relaxation times  in the system (e.g. leak rate and synaptic decay rate).  Here, for mathematical purposes it is easier to take the limit $s \to -\infty$.

Since $g_{kj}(t)$ depends on the raster plot up to time $t$, via the spiking times $\tjro$
this limit makes only sense when taking it ``conditionally'' to a prescribed raster plot $\omega$.
In other words, one can know the value of the conductances $g_{kj}$ at time $t$ only if the
past spike times of the network are known. We write $g_{kj}(t,\omega)$ from now on to make this dependence
explicit.

We set 

\beq\label{alpha_kj}
\akj{t,\omega}= \lim_{s \to -\infty} \sum_{ \Setc{r}{s \leq \tjro < t}} \akj{t \, - \,  \tjro}
\equiv \sum_{ \Setc{r}{\tjro < t}} \akj{t \, - \,  \tjro},
\eeq
with the convention that $\sum_{\emptyset}=0$ so that $\akj{t,\Omega_0}=0$
(recall that $\Omega_0$ is the raster such that no neuron ever fires).
The limit (\ref{alpha_kj}) exists from (\ref{int_alpha}). 

\ssu{Noise}\label{Snoise}

We allow, in the definition  of the  current
$I_k(t)$ in eq. (\ref{diff_gen}) the possibility of having a stochastic term corresponding
to noise so that:
\beq\label{Inoise}
I_k(t) \, =  \, i^{(ext)}_k(t) \, + \, \sigma_B \xi_k(t),
\eeq
where $i^{(ext)}_k(t)$ is a deterministic external current and $\xi_k(t)$ a noise term whose amplitude is controlled by $\sigma_B>0$. The model affords an extension
where $\sigma_B$ depends on $k$ but this extension is straightforward and we do not develop it here. 
The noise term  can be interpreted as the random variation
in the ionic flux of charges crossing the membrane per unit time at the post synaptic button, upon opening of ionic channels due to the binding of neurotransmitter.

We assume that $\xi_k(t)$ is a white noise, $\xi_k(t)=\frac{dB_k}{dt}$ where
 $dB_k(t)$ is a Wiener process, so that $dB(t)=\pare{dB_k(t)}_{k=1}^N$ is a $N$-dimensional Wiener process.
Call $P$ the noise probability distribution  and $\Exp{}$ the expectation under $P$. Then, by definition, $\Exp{dB_k(t)} = 0$, $\forall k =1 \dots N, \, t \in \setR$,
and $\Exp{dB_k(s) \, dB_l(t)} = \delta_{kl} \,\delta(t-s) dt$ where $\delta_{kl} =1$ if $l=k$, 
$l,k =1 \dots N$ and $\delta(t-s)$ is the Dirac distribution.

\ssu{Differential equation for the Integrate regime of gIF}

Summarizing, we write eq. (\ref{diff_gen}) in the form:
\beq\label{DNN}
C_k \, \frac{dV_k}{dt}+\gk{t,\omega} V_k=i_k(t,\omega),
\eeq
where:
\beq\label{gkt}
\gk{t,\omega} = \gLk + \sum_{j=1}^N G_{kj}\akj{t,\omega}.
\eeq
This is the more general conductance form considered in this paper.

Moreover, :
\beq\label{ik}
i_k(t,\omega)=\gLk \, E_L 
+  \, \sum_{j=1}^N  W_{kj} \, \akj{t,\omega}
\, + \, i^{(ext)}_k(t) \, + \, \sigma_B \xi_k(t),
\eeq 
where $W_{kj}$ is  the synaptic weight:
$$\left\{
\baR{ccc}
 W_{kj}=E^+G_{kj}, \quad &\mbox{if}&  \quad j \in \cE, \\
 W_{kj}=E^-G_{kj}, \quad &\mbox{if}&  \quad j \in \cI. 
\eaR
\right. 
$$
These equations hold when the membrane potential is below the threshold (Integrate regime).\\

Therefore, gIF models constitute rather complex
dynamical systems: the vector field (r.h.s) 
of the differential equation (\ref{DNN})
depends on an auxiliary ``variable'' which is the past spike sequence $\sif{\ent{t}}$
and to define properly the evolution of $V_k$ from time $t$ to later times one needs to know
the spikes arising before $t$. This is precisely
what makes gIF models more interesting than LIF. The definition of conductances
introduces long term \textit{memory} effects. 

 IF models implement a reset mechanism on the membrane potential:
If neuron $k$ has been reset between $s$ and $t$, say at time  $\tau$, then $V_k(t)$ depends only on 
$V_k(\tau)$
and not on previous values, as in (\ref{Fire}). But,
in gIF model, contrarily to LIF, there is also a dependence in the past
via the conductance  and this dependence \textit{is not}
erased by the reset. That's why we have to consider a system with
infinite memory.

\ssu{The parameters space}

The stochastic dynamical system (\ref{DNN}) depends on a huge set of parameters:
the membrane capacities $C_k, \, k=1 \dots N$, the threshold $\theta$, the reversal
potentials $E_L$, $E^{+},E^{-}$, the leak conductance $g_L$; the maximal synaptic conductances $G_{kj}, \, k,j=1 \dots N$
which define the neural network topology; the characteristic times $\tau_{kj},  \, k,j=1 \dots N$ of synaptic
responses decay; the noise amplitude $\sigma_B$ and, additionally, the parameters defining the external current $i_k^{(ext)}$.
Although some parameters can be fixed from biology, such as $C_k$, the reversal potentials, $\tau_{kj}$, ... 
some others such as the $G_{kj}$'s must be allowed to vary freely in order to leave open the possibility of modelling
very different neural networks structures. 

In this paper we are not interested in describing properties arising for specific values of those parameters,
but instead in generic properties that hold on sets of parameters. More specifically, we denote the list
of all parameters $\pare{\pare{C_k}_{k=1}^N, \, E_L, \, E^+, \, E^-, \, \pare{G_{kj}}_{k,j=1}^N, \dots}$ 
by the symbol $\gamma$. This is a vector in $\bbbr^K$ where $K$ is the total number of parameters.
In this paper, we assume that $\gamma$ belongs to a bounded subset $\cH \subset \bbbr^K$. Basically, 
we want to avoid situations where some parameters become infinite, which would be unphysical. 
So the limits of $\cH$ are the limits imposed by biophysics.  Additionally,
we  assume that $\sr>0$ and $\sigma_B>0$.  Together with
physical constraints such as ``conductances are positive'', these are the only assumption made 
in parameters. All mathematical results stated in the paper old for any $\gamma \in \cH$.  

\su{gIF model-dynamics  for a fixed raster}\label{SgIF_fixed_rast}

We assume that the raster $\omega$ is fixed, namely the spike history is given.  Then, it is possible to integrate the equation (\ref{DNN})
(Integrate regime) and to obtain explicitly
the value of the membrane potential of a neuron at time $t$,
given the membrane potential value at time $s$.
Additionally, the reset condition (\ref{Fire}) has the consequence of removing the dependence of neuron $k$ on the past anterior to $\tko$.
 
\ssu{Integrate regime}

For $t_1 \leq t_2$, $t_1,t_2 \in \setR$, set:
\beq\label{Gamma}
\Gamma_k(t_1,t_2,\omega)=e^{-\frac{1}{C_k}\int_{t_1}^{t_2}\gk{u,\omega} \, du}.
\eeq
We have:
$$\Gamma_k(t_1,t_1,\omega)=\Gamma_k(t_2,t_2,\omega)=1,$$
and:
$$
\frac{\partial \Gamma_k(t_1,t_2,\omega)}{\partial t_1}=\frac{\gk{t_1,\omega}}{C_k} \, \Gamma_k(t_1,t_2,\omega).
$$
Fix two times $s < t$ and  assume that for neuron $k$,  $V_k(u) <\theta$, $s \leq u \leq t$
so that the membrane potential $V_k$ obeys (\ref{DNN}).
Then:
$$\frac{\partial }{\partial t_1}\left[\Gamma_k(t_1,t_2,\omega) \, V_k(t_1)\right]=
\Gamma_k(t_1,t_2,\omega)
\left[\frac{dV_k}{dt_1}+ \frac{\gk{t_1,\omega}}{C_k} \, V_k(t_1)\right]
=\Gamma_k(t_1,t_2,\omega) \, \frac{i_k(t_1,\omega)}{C_k}.$$

We have then,
integrating the previous equation with respect to $t_1$ between $s$
and $t$, and setting $t_2=t$:


$$
V_k(t)=\Gamma_k(s,t,\omega) \, V_k(s) \,
+
\, \frac{1}{C_k} \, \int_{s}^{t} \Gamma_k(t_1,t,\omega) \,  i_k(t_1,\omega) \, dt_1.
$$

\nid This equation gives the variation of membrane potential during a period of rest (no spike) of the neuron.
Note however that this neuron can still receive spikes from the other neurons via the update of conductances
(made explicit in the previous equation by the dependence in the raster plot $\omega$).\\

The term $\Gamma_k(s,t,\omega)$ given by (\ref{Gamma}) is an effective leak
between $s,t$. In the Leaky Integrate and Fire model it would have been equal to $e^{-\int_{s}^{t}\frac{1}{\tau_L} \, dt_1} \, 
= e^{- \frac{t-s}{\tau_L}}$. The term:
%
$$
\Vksto \, \deq \, \frac{1}{C_k}\int_{s}^{t} \Gamma_k(t_1,t,\omega) \,  i_k(t_1,\omega) \, dt_1,
$$
%
has the dimension of a voltage. It
corresponds to the integration of the total current between $s$ and $t$ weighted by the effective leak term $\Gamma_k(t_1,t,\omega)$. It
decomposes as
$$
\Vksto=
\Vkosyn{s}{t} \, + \, \Vkoext{s}{t} \,
+ \, \VkoB{s}{t},$$
where,
\beq\label{Jsyn}
\Vkosyn{s}{t} \, = \, 
 \frac{1}{C_k} \sum_{j=1}^N   W_{kj} \, \int_{s}^{t} \Gamma_k(t_1,t,\omega)  \, \akj{t_1,\omega} dt_1 ,
\eeq
is the synaptic contribution. Moreover,
%
$$
\Vkoext{s}{t} \, = \,
\frac{E_L}{\tLk} \int_{s}^{t} \Gamma_k(t_1,t,\omega)   dt_1
+  \frac{1}{C_k} \, \int_{s}^{t} i^{(ext)}_k(t_1) \, \Gamma_k(t_1,t,\omega)  dt_1,
$$
%
 where we set:
\beq\label{taulk}
\tLk \deq \frac{C_k}{\gLk}, 
\eeq
the characteristic leak time of neuron $k$. We have included the leak reversal potential term in this ``external'' term for convenience. Therefore, even if there is no external current this term is nevertheless non zero.

The sum of the synaptic and external terms gives the deterministic contribution in the membrane potential. We note:
%
$$
\Vkodet{s}{t}  =  \Vkosyn{s}{t} +\Vkoext{s}{t}.
$$

Finally,
\beq\label{VB}
\VkoB{s}{t} \, \deq \,
\frac{\sigma_B}{C_k} \int_{s}^{t}  \Gamma_k(t_1,t,\omega)  \xi_k(t_1)dt_1,
=
\frac{\sigma_B}{C_k} \int_{s}^{t}  \Gamma_k(t_1,t,\omega)  dB_k(t_1),
\eeq
is a noise term. This is a Gaussian process with mean $0$ and variance:

\beq\label{sigma_k_B}
\left(\frac{\sigma_B}{C_k}\right)^2 \, \Exp{\left(\int_{s}^{t}  \Gamma_k(t_1,t,\omega)  dB_k(t_1)\right)^2}
= \left(\frac{\sigma_B}{C_k}\right)^2 \,
\int_{s}^{t} \Gamma^2_k(t_1,t,\omega)  \, dt_1.
\eeq
The square root of  this quantity has the dimension of a voltage.\\

As a final result, for a fixed $\omega$, the 
variation of membrane potential during a period of rest (no spike) of neuron $k$ between $s$ and $t$ reads (sub-threshold oscillations):

\beq \label{restV}
V_k(t)= \Gamma_k(s,t,\omega)V_k(s)+\Vkodet{s}{t} + \VkoB{s}{t}.
\eeq

\ssu{Reset}\label{SVks}

In eq. (\ref{Fire}), as in all IF models that we know,
the reset of the membrane potential has the effect of removing
the dependence of $V_k$ on its past since $V_k(\ent{t+\tau_{sep}})$ is replaced
by $\Vr$.
Hence, reset removes the dependence in the initial condition $V_k(s)$ in 
(\ref{restV})
provided that neuron $k$ fires between $s$ and $t$ in the raster $\omega$.
As a consequence, eq.  (\ref{restV}) holds,
from  the ``last reset time''
introduced in section \ref{LastReset}  up to time $t$. Then, eq. (\ref{restV})  reads 
\beq\label{Vkreset}
V_k(t)= \Vkodet{\tko}{t} +  \Vkonoise{\tko}{t}.
\eeq
where:
\beq\label{Vknoise}
\Vkonoise{\tko}{t} = \Gamma_k(\tko,t,\omega)\Vr + \VkoB{\tko}{t},
\eeq
is a Gaussian process with mean zero and variance:
\beq\label{sigma_k}
\skd{\tko,t,\omega}
= \Gamma^2_k(\tko,t,\omega) \, \sdr \, + \, \left(\frac{\sigma_B}{C_k}\right)^2 \,
\int_{\tko}^{t} \Gamma^2_k(t_1,t,\omega)  \, dt_1.
\eeq

Although the reset condition may look as
a simplification it is in fact a source of complications
on mathematical grounds. As discussed in section \ref{Sgencond} several proofs are easier if we do not reset memory and instead take the limit $s \to -\infty$ in (\ref{restV}). However, since IF models are widespread in the neuroscience literature we have preferred to give the more general proofs and then discuss their adaptation to simpler cases.

\su{Useful bounds} \label{Sbounds}

We now prove several bounds used throughout the paper.

\ssu{Bounds on the conductance}

From (\ref{int_alpha}), and since $\akj{t} \geq 0$:
\beq\label{bounds_alpha_kj}
0=\akj{t,\Omega_0}  \leq \akj{t,\omega} \, = \, \sum_{ \Setc{r}{\tjro < t}} \akj{t \, - \,  \tjro}
\leq \sum_{r < t} \akj{t-r} = \akj{t,\Omega_1}
\leq  \alpha^+.
\eeq

Therefore, 

\beq\label{bounds_gk}
\gLk = g_{k}(t,\Omega_0)   \leq g_{k}(t,\omega) \leq 
g_{k}(t,\Omega_1)  \deq \gmk \leq \gLk  + \alpha^+ \, \sum_{j=1}^N \, G_{kj} ,
\eeq
so that  the conductance is uniformly bounded in $t$ and $\omega$. The minimal
conductance is attained when no neuron fires ever so that $\Omega_0$ is the ``lowest conductance state''.
On the opposite the maximal conductance
is reached when all neurons fire all the time so that  $\Omega_1$ is the ``highest conductance state''.
To simplify notations we note $\tmk=\frac{C_k}{\gmk}$. This is the
minimal relaxation time scale for neuron $k$ while $\tLk=\frac{C_k}{\gLk}$ is the maximal relaxation time. 
\beq\label{ineq_relax}
\tmk=\frac{C_k}{\gmk} \leq \tLk=\frac{C_k}{\gLk}.
\eeq

\ssu{Bounds on membrane potential}  

Now, from (\ref{Gamma}), we have, for $s < t$:
\beq\label{bounds_Gammak}
0 \leq  \Gamma_k(s,t,\Omega_1)=
e^{- \frac{t-s}{\tmk}} \leq \Gamma_k(s,t,\omega) \leq
\Gamma_k(s,t,\Omega_0)= e^{- \frac{t-s}{\tLk}} < 1.
\eeq
As a consequence, $\Gamma_k(s,t,\omega) \to 0$ exponentially fast as $s \to -\infty$. \\

Moreover, 
\beq\label{boundsalphaGamma}
0 \leq \int_{s}^{t} \Gamma_k(t_1,t,\omega) \,  \akj{t_1,\omega} dt_1 \leq
\alpha^+ \, \int_{s}^{t} e^{- \frac{t-t_1}{\tLk}} dt_1
= \alpha^+ \, \tLk \left(1 \, -  \, e^{-\frac{t-s}{\tLk}} \right) \leq \alpha^+ \, \tLk,
\eeq
so that:
\beq\label{bounds_Vsyn}
\frac{\alpha^+}{\gLk} \, \sum_{j \in \cI}  W_{kj} 
\leq
\Vkosyn{s}{t}
\leq
\frac{\alpha^+}{\gLk} \, \sum_{j \in \cE}  W_{kj}.
\eeq
Thus, $\Vkosyn{s}{t}$ is uniformly bounded in $s,t$.

\medskip

Establishing similar bounds for $\Vkoext{s}{t}$ requires the assumption that
$A  \leq i^{(ext)}_k(t) \leq B$,  but obtaining tighter bounds requires
additionally the knowledge of the sign of $ \frac{A}{C_k}+\frac{E_L}{\tLk}$
and of  $\frac{B}{C_k}+\frac{E_L}{\tLk}$. Here, we have only to consider that:
%
$$
\abs{i^{(ext)}_k(t)} \leq i^+.
$$
%
In this case,
$$
\abs{\Vkoext{s}{t}} \, = \,
\abs{\frac{E_L}{\tLk} \int_{s}^{t} \Gamma_k(t_1,t,\omega)   dt_1
+  \frac{1}{C_k} \, \int_{s}^{t} i^{(ext)}_k(t_1) \Gamma_k(t_1,t,\omega)  dt_1}
$$
$$
\leq
\bra{\frac{\abs{E_L}}{\tLk}
+   \frac{i^+}{C_k}} \, \int_{s}^{t}  \Gamma_k(t_1,t,\omega)  dt_1
\leq
\bra{\frac{\abs{E_L}}{\tLk}
+   \frac{i^+}{C_k}} \, \int_{s}^{t}  e^{-\frac{t-t_1}{\tLk}}  dt_1,$$
so that:
\beq\label{bounds_Vext}
\abs{\Vkoext{s}{t}}
\leq
\pare{\abs{E_L}+\frac{i^+}{\gLk}}
\left(1 - e^{-\frac{t-s}{\tLk}}\right)
\leq \abs{E_L}+\frac{i^+}{\gLk}.
\eeq
Consequently:
\bp\label{Lbounds_Vkdet}
\beq\label{bounds_Vkdet}
V_k^- \deq \frac{\alpha^+}{\gLk} \, \sum_{j \in \cI}  W_{kj} 
-\pare{\abs{E_L}+\frac{i^+}{\gLk}}
<
\Vkdet{s,t,\omega}
<
V_k^+ \deq \frac{\alpha^+}{\gLk} \, \sum_{j \in \cE}  W_{kj}
+ \abs{E_L}+\frac{i^+}{\gLk},
\eeq
\ep
which provides uniform bounds in $s,t,\omega$ for the deterministic
part of the membrane potential.

\ssu{Bounds on the noise variance} 

Let us now consider the stochastic part $\Vknoise{\tko,t,\omega}$.
It has zero mean and its variance (\ref{sigma_k}) obeys the bounds:
$$
e^{-2 \, \frac{t-\tko}{\tmk}} \, \sdr \, + \, 
\frac{\tmk}{2} \, \left(\frac{\sigma_B}{C_k}\right)^2 
\, \pare{1 \, - \, e^{-2 \, \frac{t-\tko}{\tmk}}}
\leq 
\skd{\tko,t,\omega}$$
$$
\leq
e^{-2 \, \frac{t-\tko}{\tLk}} \, \sdr \, + \, 
\frac{\tLk}{2} \, \left(\frac{\sigma_B}{C_k}\right)^2 
\, \pare{1 \, - \, e^{-2 \, \frac{t-\tko}{\tLk}}}.
$$

If $\sdr < \frac{\tmk}{2} \, \left(\frac{\sigma_B}{C_k}\right)^2$
the left hand side is an increasing function of $u=t-\tko \geq 0$ so that
the minimum, $\sdr$ is reached for $u=0$ while the maximum is reached for
$u=+\infty$ and is $\frac{\tmk}{2} \, \left(\frac{\sigma_B}{C_k}\right)^2$.
The opposite holds if $\sdr \geq \frac{\tmk}{2} \, \left(\frac{\sigma_B}{C_k}\right)^2$. The same argument holds \textit{mutatis mutandis} for the right hand side.
We set:
\beq\label{sigma_pm}
\sigma_k^- \deq \min\pare{\frac{\sigma_B}{C_k} \, \sqrt{\frac{\tmk}{2}},\sr}; \quad
\sigma_k^+ \deq \max\pare{\frac{\sigma_B}{C_k} \, \sqrt{\frac{\tLk}{2}},\sr}
\eeq
so that:
\bp\label{Pbounds_sigma_k}
\beq\label{bounds_sigma_k}
0 < \sigma_k^- 
\leq \sigma_k(\tko,t,\omega)
\leq \sigma_k^+
< +\infty.
\eeq
\ep

\ssu{The limit $\tko \to -\infty$}

For fixed $s$ and $t$ there are infinitely many rasters such
that $\tko < s$ (we remind that rasters are infinite sequences).
One may argue
that taking the difference $t-s$ sufficiently large the probability of such
sequences should vanish. It is indeed possible to show (section \ref{SPsilent})
that this
probability vanishes exponentially fast with $t-s$, meaning unfortunately that it is  \textit{positive}
whatever $t-s$. So we have to consider cases where $\tko$ can go arbitrary far in past
(this is also a key toward an extension of the present analysis to more general conductance-based models as discussed in section  \ref{Sgencond}). Therefore,
we have to check that the quantities introduced in the previous sections are
well defined as $\tko \to -\infty$. 

Fix $s$ real. For all $\omega$ such that $\tko \leq s$ - this condition ensuring that $k$ does not fire
between $s$ and $t$ - we have, from (\ref{bounds_alpha_kj}), (\ref{bounds_Gammak}), 
$0 \leq \Gamma_k(t_1,t,\omega)  \akj{t_1,\omega} \leq \alpha^+ \, e^{- \frac{t-t_1}{\tLk}}$,
$\forall t_1 \in [s,t]$.
Now, since $\lim_{s \to -\infty} \int_s^t  e^{- \frac{t-t_1}{\tLk}} dt_1 = \tLk$ exists,
the limit 
$$\lim_{\tko \to -\infty} \Vkosyn{\tko}{t}= \frac{1}{C_k} \sum_{j=1}^N   W_{kj} \, \int_{-\infty}^t \Gamma_k(t_1,t,\omega)  \akj{t_1,\omega} dt_1.$$
exists as well. The same holds for the external term $\Vkoext{\tko}{t}$.
%
%
 
Finally, since $\Gamma_k(\tko,t,\omega) \to 0$
as $\tko \to -\infty$ the noise term (\ref{Vknoise}) becomes in the limit:
%
$$
\lim_{\tko \to -\infty} \Vkonoise{\tko}{t} \, = \,
\frac{\sigma_B}{C_k} \int_{-\infty}^{t}  \Gamma_k(t_1,t,\omega)  dB_k(t_1),
$$
%
which is a Gaussian process with mean $0$ and a variance 
$\pare{\frac{\sigma_B}{C_k}}^2 \int_{-\infty}^{t}  \Gamma^2_k(t_1,t,\omega)  dt_1$
which obeys the bounds (\ref{bounds_sigma_k}).

\su{Continuity with respect to a raster.} \label{Scont}

\ssu{Definition}
Due to the particular structure of gIF models we have seen that
the membrane potential at time $t$ is both a function of $t$
and of the full sequence of past spikes $\sif{\ent{t}}$. One expects 
however the dependence with respect to the past spikes to decay as
those spikes are more distant in the past. This issue is related
to a notion of continuity with respect to a raster that we now characterize.

\bdf Let $m$ be a positive integer.The $m$-\textit{variation} of a function $f(t,\omega) \equiv f(t,\sif{\ent{t}})$ is:
\beq\label{varm}
\vm{f(t,.)}=\sup\Set{ \, | \,f(t,\omega)-f(t,\omega')\, | \,: \omega \eg{m,\ent{t}} \omega' }.
\eeq
\edf
where the definition of $\eg{m,\ent{t}}$ is given in eq. (\ref{eg}).
Hence, this notion characterizes the maximal variation of $f(t,.)$ on the set of spikes
identical from time $\ent{t}-m$ to time $\ent{t}$ (cylinder set). It implements the fact that one may truncate
the spike history to time $\ent{t}-m$ and make an error which is at most $\vm{f(t,.)}$.

\bdf
The function $f(t,\omega)$ is \textit{continuous} if $\vm{f(t,.)} \to 0$ as $m \to +\infty.$
\edf

An additional information is provided by the  convergence rate to $0$ with $m$.
The faster this convergence the smaller the error made when replacing an infinite
raster by a spike block on a finite time horizon. 

\ssu{Continuity of conductances} 

\bp\label{PContCond}
The conductance $\gk{t,\omega}$ is continuous in $\omega$, for all $t$, for all $k=1 \dots N$.
\ep

\bpr
Fix $k=1 \dots N$, $t \in \setR$, $m > 0$ integer. We have, for 
$\omega \eg{m,\ent{t}} \omega'$:
$$
\abs{\akj{t,\omega} \, - \, \akj{t,\omega'}}
= \abs{
\sum_{ \Setc{r}{\tjro < t}} \akj{t \, - \,  \tjro}
\, - \,
\sum_{ \Setc{r'}{\tj{r'}{\omega'} < t}} \akj{t \, - \,  \tj{r'}{\omega'}}
}$$
$$
= \abs{
\sum_{ \Setc{r}{\tjro < t-m}} \akj{t \, - \,  \tjro}
\quad - \,
\sum_{ \Setc{r'}{\tj{r'}{\omega'} < t-m}} \akj{t \, - \,  \tj{r'}{\omega'}}
},
$$
since the set of firing times 
$\Set{t-m \leq \tjro < t }$,  
$\Set{t-m \leq  \, \tj{n'}{\omega'} < t }$ are identical by hypothesis.
So, since $\akj{x} \geq 0$,
$$
\abs{\akj{t,\omega} \quad - \, \akj{t,\omega'}}
$$
$$
\leq
\sum_{ \Setc{r}{\tjro < t-m}} \akj{t \quad - \,  \tjro}
\, + \,
\sum_{ \Setc{r'}{\tj{r'}{\omega'} < t-m}} \akj{t \quad - \,  \tj{r'}{\omega'}}
$$
$$
\leq
2 \, \sum_{r < t-m} \akj{t-r}.
$$
Therefore, as $m\to +\infty$, from (\ref{lim_alpha}) and setting $M=t-m$, :
%
$$\vm{\akj{t,.}} \leq 2 \,\, \sum_{r < M} \akj{t-r} \preceq 2 \,  \, P_d\pare{\frac{m}{\tau_{kj}}} \, 
e^{-\frac{m}{\tau_{kj}}},
$$
which converges to $0$ as $m \to +\infty$. 

Therefore, from (\ref{gkt}) $\gk{t,\omega}$ is continuous with a variation
%
$$
\vm{\gk{t,.}} \preceq 2 \, \sum_{j=1}^N G_{kj}  \, P_d\pare{\frac{m}{\tau_{kj}}} \, 
e^{-\frac{m}{\tau_{kj}}},
$$
which converges exponentially fast to $0$ as $m \to +\infty$.
\epr

\ssu{Continuity of the membrane potentials}

\bp\label{PContV}
The deterministic part of the membrane potential, $\Vkdet{\tko,t,\omega}$,
is continuous and its $m$-variation  decays exponentially fast with $m$.
\ep

\bpr
In the proof, we shall establish precise upper bounds for the variation of 
$\Vksyn{\tau_k(t,.),t,.}$, $\Vkext{\tau_k(t,.),t,.}$ since they are used later on for the proof of uniqueness of a Gibbs measure (section \ref{SGibbs}).
From the previous result it is easy to show that, for all $\omega \eg{m,\ent{t}} \omega'$, $t_1 \leq t_2 \leq t$:
$$
\abs{\Gamma_k(t_1,t_2,\omega) \, - \, \Gamma_k(t_1,t_2,\omega')}
\leq \Gamma_k(t_1,t_2,\omega) \pare{e^{\frac{\vm{g_k}}{C_k}(t_2 - t_1)}-1}, $$
Therefore, from (\ref{bounds_Gammak}),
%
%
$$
\vm{\Gamma_k(t_1,t_2, .)} \leq
e^{-\frac{t_2 - t_1}{\tLk}}\pare{e^{\frac{\vm{g_k}}{C_k}(t_2 - t_1)}-1} 
$$
%
and $\Gamma_k(t_1,t_2,\omega)$ is continuous in $\omega$.\\

Now the product $\Gamma_k(t_1,t_2,\omega) \, \akj{t_1,\omega}$ is continuous
as a product of continuous functions. Moreover:
$$\vm{\Gamma_k(t_1,t_2, .)  \akj{t_1,.}} \leq 
\sup_{\omega \in X}\Gamma_k(t_1,t_2,\omega)\, \vm{\akj{t_1,.}}
+
\sup_{\omega \in X}\akj{t_1,\omega} \vm{\Gamma_k(t_1,t_2,.)}$$
$$
=e^{-\frac{t_2 - t_1}{\tLk}} \bra{\vm{\akj{t_1,.}}
+
\pare{ e^{\frac{\vm{g_k}}{C_k}(t_2 - t_1)}-1} \akj{t_1,\Omega_1}}
,$$
so that:
$$
\vm{\Gamma_k(t_1,t_2, .)  \akj{t_1,.}}
< e^{-\frac{t_2 - t_1}{\tLk}} \bra{ 2 \,  \, P_d\pare{\frac{m}{\tau_{kj}}} \, e^{-\frac{m}{\tau_{kj}}}
+
\pare{ e^{\frac{\vm{g_k}}{C_k}(t_2 - t_1)}-1} \alpha^+}.
$$

Since, as $m \to +\infty$:
$$
e^{\frac{\vm{g_k}}{C_k}(t_2 - t_1)}-1
\sim 
\frac{\vm{g_k}}{C_k}(t_2 - t_1)$$
$$
\preceq \frac{2(t_2 - t_1)}{C_k} \, \sum_{j'=1}^N G_{kj'}  P_d\pare{\frac{m}{\tau_{kj'}}} \, 
e^{-\frac{m}{\tau_{kj'}}},$$
we have,
%
$$
\vm{\Gamma_k(t_1,t_2, .)  \akj{t_1,.}}
\preceq 2 \, e^{-\frac{t_2 - t_1}{\tLk}}\bra{  \, P_d\pare{\frac{m}{\tau_{kj}}} \, e^{-\frac{m}{\tau_{kj}}}
+
\frac{ \alpha^+ (t_2 - t_1)}{C_k} \, \D{\sum_{j'=1}^N} G_{kj'}  \, P_d\pare{\frac{m}{\tau_{kj'}}} \, 
e^{-\frac{m}{\tau_{kj'}}}},
$$
%
which converges to $0$ as $m \to +\infty$.\\

Let us show the continuity of $\Vksyn{\tau_k(t,.),t,.}$. 
We have, from (\ref{Jsyn}),
$$
\abs{\Vksyn{\tko,t,\omega}-\Vksyn{\tkop,t,\omega'}} \, 
\leq \,$$
$$ 
 \frac{1}{C_k} \sum_{j=1}^N   \abs{W_{kj}} \, 
\abs{
\int_{\tko}^{t} \Gamma_k(t_1,t,\omega)  \akj{t_1,\omega} dt_1
-
\int_{\tkop}^{t} \Gamma_k(t_1,t,\omega')  \akj{t_1,\omega'} dt_1
}.
$$
The following inequality is used at several places in the paper.
For a $t_1$-integrable function $f(t_1,t,\omega)$, we have:
\beq\label{Ineg_f}
\abs{
\int_{\tko}^{t} f(t_1,t,\omega) dt_1
-
\int_{\tkop}^{t} f(t_1,t,\omega') dt_1
}
\leq
\int_{\tko}^{t} 
\abs{
f(t_1,t,\omega
-
f(t_1,t,\omega')
} dt_1
+
\abs{
\int_{\tko}^{\tkop} f(t_1,t,\omega') dt_1
}.
\eeq

Here it gives,  for $t_1 \leq t$:
$$\abs{
\int_{\tko}^{t} \Gamma_k(t_1,t,\omega)  \akj{t_1,\omega} dt_1
-
\int_{\tkop}^{t} \Gamma_k(t_1,t,\omega')  \akj{t_1,\omega'} dt_1
}$$
$$\leq
\int_{\tko}^{t} 
\abs{
\Gamma_k(t_1,t,\omega)  \akj{t_1,\omega}
-
\Gamma_k(t_1,t,\omega')  \akj{t_1,\omega'}
} dt_1
+
\abs{
\int_{\tko}^{\tkop} \Gamma_k(t_1,t,\omega')  \akj{t_1,\omega'} dt_1
}.
$$

For the first term, we have,
$$
\int_{\tko}^{t} 
\abs{
\Gamma_k(t_1,t,\omega)  \akj{t_1,\omega}
-
\Gamma_k(t_1,t,\omega')  \akj{t_1,\omega'}
} dt_1 
$$
$$
\leq 
\int_{\tko}^{t} 
\vm{\Gamma_k(t_1,t,.) \, \akj{t_1,.}}
 dt_1
\leq \int_{-\infty}^{t} 
\vm{\Gamma_k(t_1,t,.) \, \akj{t_1,.}} dt_1
$$
$$
\preceq 
\int_{-\infty}^{t} 
2 \, e^{-\frac{t - t_1}{\tLk}}\bra{  \, P_d\pare{\frac{m}{\tau_{kj}}} \, e^{-\frac{m}{\tau_{kj}}}
+
\frac{ \alpha^+ (t - t_1)}{C_k} \, \sum_{j'=1}^N G_{kj'}  \, P_d\pare{\frac{m}{\tau_{kj'}}} \, 
e^{-\frac{m}{\tau_{kj'}}}} dt_1,
$$
$$
= 
2  \, P_d\pare{\frac{m}{\tau_{kj}}} \, e^{-\frac{m}{\tau_{kj}}}
\int_{-\infty}^{t} 
 \, e^{-\frac{t - t_1}{\tLk}} dt_1
\, + \, 
\frac{ 2 \, \alpha^+}{C_k} \, 
\sum_{j'=1}^N G_{kj'}  \, P_d\pare{\frac{m}{\tau_{kj'}}} \, 
e^{-\frac{m}{\tau_{kj'}}} \int_{-\infty}^{t} 
 (t - t_1) \, e^{-\frac{t - t_1}{\tLk}} dt_1,
$$
$$
= 2 \, \tLk \bra{
  \, P_d\pare{\frac{m}{\tau_{kj}}} \, e^{-\frac{m}{\tau_{kj}}}
\, + \, 
\frac{\alpha^+ \, \tLk}{C_k} \, 
\sum_{j'=1}^N G_{kj'}  \, P_d\pare{\frac{m}{\tau_{kj'}}} \, 
e^{-\frac{m}{\tau_{kj'}}} 
}.$$

Let us now consider the second term. If $\tko \geq t-m$ or $\tkop \geq t-m$, 
then $\tko=\tkop$ and this term vanishes.
Therefore, the supremum in the definition of $\vm{\Vksyn{\tau_k(t,.),t,.}}$ is attained
if $\tko < t-m$ and $\tkop < t-m$.  We may assume, without loss of generality, 
that $\tkop \geq \tko$. Then, from (\ref{boundsalphaGamma}),
$$
\int_{\tko}^{\tkop} \Gamma_k(t_1,t,\omega')  \akj{t_1,\omega'} dt_1
<
\alpha^+ \int_{\tko}^{\tkop}e^{-\frac{t - t_1}{\tLk}} dt_1$$
$$ =
\alpha^+ \tLk \, e^{-\frac{t - \tkop}{\tLk}} \pare{1 \, - \, e^{-\frac{\tkop - \tko}{\tLk}} }
\leq \alpha^+ \tLk \, e^{-\frac{t - \tkop}{\tLk}} \leq  \alpha^+ \tLk e^{-\frac{m}{\tLk}}.
$$

So, we have, for the variation of $\Vksyn{\tau_k(t,.),t,.}$, using (\ref{taulk}):
$$
\vm{\Vksyn{\tau_k(t,.),t,.}} \preceq  $$
$$
\frac{1}{\gLk} \sum_{j=1}^N   \abs{W_{kj}} \,
\bra{
 2 \,  \pare{
  \, P_d\pare{\frac{m}{\tau_{kj}}} \, e^{-\frac{m}{\tau_{kj}}}
\, + \, 
\frac{\alpha^+}{\gLk} \, 
\sum_{j'=1}^N G_{kj'}  \, P_d\pare{\frac{m}{\tau_{kj'}}} \, 
e^{-\frac{m}{\tau_{kj'}}} 
}
\, + \, 
\alpha^+  e^{-\frac{m}{\tLk}}
},
$$
so that finally,
\beq\label{VarVksyn}
\vm{\Vksyn{\tau_k(t,.),t,.}} \preceq
\sum_{j=1}^N \cst{A}{kj}{syn} \, P_d\pare{\frac{m}{\tau_{kj}}} \, e^{-\frac{m}{\tau_{kj}}}
\, + \, \cst{B}{k}{syn} \, e^{-\frac{m}{\tLk}},
\eeq
with
\beq\label{Akjsyn}
\cst{A}{kj}{syn}=\frac{1}{\gLk} \, \pare{2 \, \abs{W_{kj}} 
\, + \,
\alpha^+ \frac{G_{kj}}{\gLk} \, 
\sum_{j'=1}^N \abs{W_{kj'}} 
},
\eeq
\beq\label{Bksyn}
\cst{B}{k}{syn} = \frac{\alpha^+}{\gLk} \, 
\sum_{j=1}^N \abs{W_{kj}},
\eeq
and $\vm{\Vksyn{\tau_k(t,.),t,.}}$ converges to $0$ exponentially fast as $m \to +\infty$.\\

Now, let us show the continuity of $\Vkext{\tko,t,\omega}$ with respect to
$\omega$. We have:
$$
\abs{\Vkext{\tko,t,\omega} - \Vkext{\tkop,t,\omega'}}$$
$$
=
\abs{
\int_{\tko}^{t}
\pare{\frac{E_L}{\tLk}  
+  \frac{i^{(ext)}_k(t_1)}{C_k}} \,   \Gamma_k(t_1,t,\omega) \, dt_1
-
\int_{\tkop}^{t}
\pare{\frac{E_L}{\tLk}  
+  \frac{i^{(ext)}_k(t_1)}{C_k}} \,   \Gamma_k(t_1,t,\omega') \, dt_1}
$$
$$
\leq
\int_{\tko}^{t}
\abs{\frac{E_L}{\tLk}  
+  \frac{i^{(ext)}_k(t_1)}{C_k}} \,   \abs{\Gamma_k(t_1,t,\omega)- \Gamma_k(t_1,t,\omega')}\, dt_1
+
\int_{\tko}^{\tkop}
\abs{\frac{E_L}{\tLk}  
+  \frac{i^{(ext)}_k(t_1)}{C_k}} \, \Gamma_k(t_1,t,\omega') \, dt_1
$$
$$
\leq
\pare{\frac{\abs{E_L}}{\tLk}  
+  \frac{i^+}{C_k}}
\pare
{\int_{\tko}^{t}
 \,   \abs{\Gamma_k(t_1,t,\omega)- \Gamma_k(t_1,t,\omega')}\, dt_1
+
\int_{\tko}^{\tkop}
  \Gamma_k(t_1,t,\omega') \, dt_1}
$$
$$
\leq
\pare{\frac{\abs{E_L}}{\tLk}  
+  \frac{i^+}{C_k}}
\pare
{\int_{\tko}^{t}
 \,   \vm{\Gamma_k(t_1,t,.)}\, dt_1
\, + \,
\int_{\tko}^{\tkop}
  e^{-\frac{t-t_1}{\tLk}} \, dt_1
}
$$
$$
\preceq 
\pare{\frac{\abs{E_L}}{\tLk}  
+  \frac{i^+}{C_k}}
\pare{ 
\frac{2 \, \tLk^2}{C_k} \, 
\sum_{j'=1}^N G_{kj'}  \, P_d\pare{\frac{m}{\tau_{kj'}}} \, e^{-\frac{m}{\tau_{kj'}}}
\,
+
\,
\tLk e^{-\frac{m}{\tLk}}
}, 
$$
where, in the last inequality, we have used that the supremum in the variation is attained for  $\tko < t-m$ and $\tkop < t-m$. Finally:
\beq\label{Var_Vkext}
\vm{\Vkext{\tau_k(t,.),t,.}}  \preceq
\sum_{j=1}^N \cst{A}{kj}{ext} \,  P_d\pare{\frac{m}{\tau_{kj}}} \, e^{-\frac{m}{\tau_{kj}}} \, + \, \cst{B}{k}{ext} \,  e^{-\frac{m}{\tLk}},
\eeq
where,
\beq\label{Akjext}
\cst{A}{kj}{ext} =2 \, \frac{G_{kj}}{\gLk}\cst{B}{k}{ext}
\eeq
\beq\label{Bkext}
\cst{B}{k}{ext} = \abs{E_L}  +  \frac{i^+}{\gLk},
\eeq
and $\Vkext{\tau_k(t,.),t,.}$ is continuous.

As a conclusion, $\Vkdet{\tau_k(t,.),t,.}$ is continuous as the sum of two continuous functions.
\epr

\ssu{Continuity of the variance of $\Vknoise{\tau_k(t,.),t,.}$}

Using the same type of arguments one can also prove that

\bp\label{PContSigma}
The variance $\sk{\tko,t,\omega}$ is continuous in $\omega$, for all $t$, for all $k=1 \dots N$.
\ep

\bpr
We have, from (\ref{sigma_k})
$$\abs{\skd{\tko,t,\omega} - \skd{\tkop,t,\omega'}}
\leq
$$
$$
\sdr \, \vm{\Gamma^2_k(\tau_k(t,.),t,.)}
\, + \, \left(\frac{\sigma_B}{C_k}\right)^2 
\abs{
\int_{\tauk{t,\omega}}^{t} \Gamma^2_k(t_1,t,\omega)  \, dt_1
-
\int_{\tauk{t,\omega'}}^{t} \Gamma^2_k(t_1,t,\omega')  \, dt_1
}.
$$

For the first term we have that the sup in $\vm{\Gamma^2_k(\tau_k(t,.),t,.)}$
is attained for $\tko,\tkop <  t-m$ and:
$$
\vm{\Gamma^2_k(\tau_k(t,.),t,.)} \leq 
2 \sup\Setc{\Gamma^2_k(\tko,t,\omega)}{\omega \, \mbox{s.t.} \, \tko <  t-m}
\leq 2 e^{-\frac{2m}{\tLk}}.
$$
For the second term we have:
$$
\abs{
\int_{\tauk{t,\omega}}^{t} \Gamma^2_k(t_1,t,\omega)  \, dt_1
-
\int_{\tauk{t,\omega'}}^{t} \Gamma^2_k(t_1,t,\omega')  \, dt_1
}
$$
$$
\leq
\int_{\tauk{t,\omega}}^{t} 
\abs{\Gamma^2_k(t_1,t,\omega) \, - \, \Gamma^2_k(t_1,t,\omega')
}\, dt_1
+
\int_{\tauk{t,\omega}}^{\tauk{t,\omega'}} 
\Gamma^2_k(t_1,t,\omega')  \, dt_1
$$
$$\leq
\int_{\tauk{t,\omega}}^{t} 
\vm{\Gamma^2_k(t_1,t,.)}\, dt_1
+
\int_{\tauk{t,\omega}}^{\tauk{t,\omega'}} 
e^{-\frac{2(t \, - \, t_1)}{\tLk}}  \, dt_1
$$
$$
\leq
\int_{-\infty}^{t} 
e^{-\frac{2(t \, - \, t_1)}{\tLk}} 
\pare{e^{\frac{2 \, \vm{\gk{t,.} \, (t-t_1)}}{C_k}} \, - \,1}\, dt_1
\, + \,
\frac{\tLk}{2} \, e^{-\frac{2 m} {\tLk}}
$$
$$
\preceq
\frac{\tLk^2}{2 \, C_k} \, \vm{\gk{t,.}} \, + \,
\frac{\tLk}{2} \, e^{-\frac{2 m} {\tLk}},
$$
so that finally:
\beq\label{Var_sigma_k}
\vm{\skd{t,.}}
\preceq
\sum_{j=1}^N  \cst{A}{kj}{\sigma} \, P_d\pare{\frac{m}{\tau_{kj}}} \, e^{-\frac{m}{\tau_{kj}}} \, + \,
\cst{C}{k}{\sigma} \, e^{-\frac{2 m} {\tLk}},
\eeq
with
%
$$
\cst{A}{kj}{\sigma}= 
\frac{G_{kj}}{\gLk} \left(\frac{\sigma_B \, \sqrt{\tLk}}{C_k}\right)^2
$$
%
$$
\cst{C}{k}{\sigma}= 
\frac{1}{2} \, \left(\frac{\sigma_B \, \sqrt{\tLk}}{C_k}\right)^2
\, + \, 2 \sdr,$$
and continuity follows.
\epr

\ssu{Remark} 
Note that the variation of all quantities considered here
is exponentially decaying with a time constant given by $\max(\tau_{kj},\tLk)$.
This is physically satisfactory: the loss of memory in the system is controlled
by the leak time and the decay of the post-synaptic potential.

\su{Statistics of  raster plots.}\label{Sstat}

\ssu{Conditional probability distribution of $V_k(t)$.}\label{SProcess}

Recall that $P$ is the joint distribution of the noise and $\Exp{}$ the expectation under $P$.
Under $P$ the membrane potential $V$ is a stochastic process whose evolution, below
the threshold, is given eq. (\ref{restV}), (\ref{Vkreset}) and above by (\ref{Fire}). 
It follows from the previous analysis that:

\bp
Conditionally to  $\sif{\ent{t}}$, 
$V(t)$ is Gaussian with mean:

$$\Expc{V_k(t)}{\sif{\ent{t}}} = \Vkdet{\tko,t,\omega}, \quad k=1 \dots N,
$$
%
and covariance:
%
$$\Cov{V_k(t),V_l(t)\, | \,\sif{\ent{t}}}= \skd{\tko,t,\omega}\delta_{kl}, \quad k,l = 1 \dots N$$
%
%
where $\skd{\tko,t,\omega}$ is given by (\ref{sigma_k}). \\

Moreover, the $V_k(t)$'s, $k=1 \dots N$,  are conditionally independent.
\ep

\bigskip

\bpr
Essentially, the proof is a direct consequence of eq. (\ref{restV}), (\ref{Vkreset}) and the Gaussian nature
of the noise $\Vknoise{\tko,t,\omega}$. The conditional independence results from the fact that:
$$
\Cov{V_k(t),\, V_l(t) \, | \,\sif{\ent{t}}} =
$$
$$
\frac{\sigma_B^2}{C_k \, C_l} \Exp{
\int_{\tko}^{t}  \Gamma_k(t_1,t,\omega)  dB_k(t_1)
\int_{\tlo}^{t}  \Gamma_l(t_2,t,\omega)  dB_l(t_2)
\, | \,\omega}
\, + \,
 \Cov{\Gamma_k(\tko,t,\omega) \, \Vr, \, \Gamma_l(\tlo,t,\omega) \, \Vr}$$
$$
= 
\frac{\sigma_B^2}{C_k \, C_l}
\int_{\tko}^{t}\int_{\tlo}^{t} 
 \Gamma_k(t_1,t,\omega)  \Gamma_l(t_2,t,\omega) 
\Exp{dB_k(t_1) dB_l(t_2)}
\, + \, \sdr \, \Gamma^2_k(\tko,t,\omega) \, \delta_{kl}
$$
$$
=
\delta_{kl}
\bra{
\left(\frac{\sigma_B}{C_k}\right)^2 
\int_{\tko}^{t}\int_{\tko}^{t} 
 \Gamma_k(t_1,t,\omega)  \Gamma_k(t_2,t,\omega) 
\delta(t_1-t_2) dt_1 dt_2 \, + \, \sdr \, \Gamma^2_k(\tko,t,\omega)}
$$
$$
= \skd{\tko,t,\omega} \delta_{kl}.
$$
\epr

\ssu{The transition probability.} 

We now compute the probability of a spiking pattern at time $n=\ent{t}$,
$\omega(n)$, given the past sequence $\sif{n-1}$. 

\bp
The probability  of $\omega(n)$ conditionally to  $\sif{n-1}$
 is given by:
\beq\label{Pt+1_cond}
\Probc{\omega(n)}{\sif{n-1}}
=\prod_{k=1}^N \Probc{\omega_k(n)}{\sif{n-1}},
\eeq

\nid with

\beq\label{Pt+1_cond_k}
 \Probc{\omega_k(n)}{\sif{n-1}}=
\omega_{k}(n) \, \pi\left(\Xk{n-1,\omega}\right)+
\left(1-\omega_{k}(n)\right) \,
\left(1-\pi\left(\Xk{n-1,\omega}\right)\right),
\eeq
where 
\beq\label{defX}
\Xk{n-1,\omega} = \frac{\theta-\Vkdet{\tau_k(n-1,\omega),n-1,\omega}}{\sk{\tau_k(n-1,\omega),n-1,\omega}},
\eeq%
and
\beq\label{pi}
\pi(x)=\frac{1}{\sqrt{2\pi}}\int_x^{+\infty} e^{-\frac{u^2}{2}}du.
\eeq
\ep

\bpr
We have, using the conditional independence of the $V_k(n)$'s:
$$
\Probc{\omega(n)}{\sif{n-1}}=
\prod_{k=1}^N
\pare{
\omega_k(n) \, 
\Probc{ V_k(n-1) \geq \theta}{\sif{n-1}}
+
(1-\omega_k(n)) \, \Probc{V_k(n-1)< \theta}{\sif{n-1}}
}.
$$
Since the $V_k(n-1)$'s are conditionally Gaussian, with mean $\Vkdet{\tau_k(n-1,\omega),n-1,\omega}$
and variance $\skd{\tau_k(n-1,\omega),n-1,\omega}$ we directly obtain (\ref{Pt+1_cond}),(\ref{Pt+1_cond_k}).

Note that since $\sk{\tau_k(n-1,\omega),n-1,\omega}$ is bounded from below 
by a positive quantity (see (\ref{bounds_sigma_k}))
the ratio $\frac{\theta - \Vkdet{\tau_k(n-1,\omega),n-1,\omega}}{\sk{\tau_k(n-1,\omega),n-1,\omega}}$ in (\ref{Pt+1_cond_k}) is defined for all $\omega \in X$.
\epr

\ssu{Chains with complete connections}

The transition probabilities (\ref{Pt+1_cond}) define
a stochastic process on the set of raster plots where
 the underlying membrane potential dynamics
is summarized in the terms
$\Vkdet{\tau_k(n-1,\omega),n-1,\omega}$ and $\sk{\tau_k(n-1,\omega),n-1,\omega}$. While
the integral defining these terms extends from $\tauk{n-1,\omega}$ to $n-1$
where $\tauk{n-1,\omega}$ can go arbitrary far in the past, the integrand
involves the conductance $g_k(n-1,\omega)$ which summarizes
an history dating back to $s = -\infty$.  
As a consequence, the probability transitions
 generate a stochastic
process with unbounded memory, thus non Markovian.
One may argue that this property is a result of our
procedure of taking the initial condition in a infinite past $s \to -\infty$,
to remove the unresolved dependency on $g_k(s)$ (section \ref{Ssyn}). 
So the alternative is either to keep $s$ finite in order to have
a Markovian process; then we have
to fix arbitrarily $g_k(s)$ and the probability  distribution of $V_k(s)$. Or
we take $s \to -\infty$, removing the initial condition, to the price of considering a non Markovian process. Actually, such processes are widely
studied in the literature under the name of ``chains with complete
connections'' \cite{chazottes:99,ledrappier:74,coelho-quas:98,bressaud-fernandez-etal:99,maillard:07}
and several important results can be used here. So we adopt the second approach of the alternative. As a by-product the knowledge of the Gibbs measure provided by this analysis allows a posteriori to fix the probability distribution of $V_k(s)$.

For the sake of completeness we give here the definition of a chain with complete connections (see \cite{maillard:07} for more details).
For $n \in \setZ$, we note $\seq{\cA}{-\infty}{n-1}$ the set of sequences $\sif{n-1}$
and $\cF_{\leq n-1}$ the related $\sigma$-algebra, while $\cF$ is the $\sigma$-algebra related with $X=\cA^\setZ$. $\cP(X,\cF)$ is the set of probability measures on $(X,\cF)$.

\bdf 
A system of transition probabilities is a family $\Set{P_n}_{n \in \setZ}$ 
of functions 
$$\Pnc{ \, }{ \, } : \cA \times \seq{\cA}{-\infty}{n-1}  \to [0, 1],$$
such that the following conditions hold for every $n \in \setZ$:

\bit
\item For every $\omega(n) \in \cA$ the function $\Pnc{\omega(n)}{.}$ is measurable with respect to $\cF_{\leq n-1}$.
\item For every $\sif{n-1} \in \seq{\cA}{-\infty}{n-1}$,
$$ \sum_{\omega(n) \in \cA} \Pnc{\omega(n)}{\sif{n-1}}=1.$$
\eit

A probability measure $\mu$ in $\cP(X,\cF)$
is consistent with a system of transition probabilities $\Set{P_n}_{n \in \setZ}$ if
for all $n \in \setZ$ and all $\cF_{\leq n}$-measurable functions $f$:
%
$$\int f\left(\seq{\omega}{-\infty}{n}\right) \mu(d\omega) = \int \sum_{\omega(n) \in \cA}
f\left(\sif{n-1} \omega(n) \right) \Pnc{\omega(n)}{\sif{n-1}} \mu(d\omega).
$$
%

 Such a measure $\mu$ is called a ``chain with
complete connections consistent with the system of transition probabilities $\Set{P_n}_{n \in \setZ}$''.
\edf

The transitions probabilities (\ref{Pt+1_cond}) constitute such a system of transitions
probabilities: the summation to $1$ is obvious while the measurability follows from the continuity of $\Pnc{\omega(n)}{\sif{n-1}}$ proved below. 
 To simplify notations we write $p(n,\omega)$ instead of $\Pnc{\omega(n)}{\sif{n-1}}$
whenever it makes no confusion. \\

\ssu{Existence of a consistent probability measure $\mu$}

In the definition above, the measure $\mu$ summarizes the statistics of spike trains from $-\infty$
to $+\infty$. Its marginals allow the characterization of finite spike blocks.
So, $\mu$ provides the characterization of spike train statistics in gIF models.
Its existence is established by a standard result in the frame of chains with complete connections
stating that a system of continuous transition probabilities on a compact space has at least one probability measure consistent with it \cite{maillard:07}. Since $\pi$ is a continuous function the continuity of
$p(n,\omega)$ with respect to $\omega$
 follows from the continuity of $\Vkdet{\tau_k(n-1,\omega),n-1,\omega}$ and the continuity of $\sk{\tau_k(n-1,\omega),n-1,\omega}$, proved in section
\ref{Scont}.

Therefore there is at least one probability measure consistent with (\ref{Pt+1_cond}).

\sssu{The Gibbs distribution.}\label{SGibbs}

A system of transition probabilities is non-null if for all $n  \in \setZ$
and all $\sif{n-1} \in \seq{\cA}{-\infty}{n-1}$, $\Probc{\omega(n)}{\sif{n-1}}>0$.
Following \cite{fernandez-maillard:05}, a chain with complete connection $\mu$ is a Gibbs measure consistent with
the system of transition probabilities $p(n,.)$ if this system is continuous and non-null. Gibbs distributions play an important role in statistical physics, as well as
ergodic theory and stochastic processes. In statistical physics they are usually derived from the maximal entropy principle \cite{jaynes:57}. Here we use them in a more general context affording to consider non-stationary processes. It turns out that the spike train statistics in gIF model is given by such a Gibbs measure. In this section we prove the main mathematical result of this paper (uniqueness of the Gibbs measure). The  consequences for spike trains characterizations are discussed in the next section.  

\bth\label{Tuniq}
For each choice of parameters $\gamma \in \cH$ 
the gIF model (\ref{DNN}) has a unique Gibbs distribution.
\enth

The proof of uniqueness is based on the following criteria due to
Fernandez and Maillard \cite{fernandez-maillard:05}.

\bp
Let:
%
$$m(p)=\inf_{n \in \setZ} \, \inf_{\omega \in \cA_{-\infty}^n} p(n,\omega),
$$
%
and \footnote{In \cite{fernandez-maillard:05} the authors use the following definition for the $n,m$ variation, which reads in our 
notations:
$$var'_m\bra{p(n,.)} = \sup\Setc{\abs{p(n,\omega) - p(n,\omega')}}{\omega, \omega' \in \cA_{-\infty}^n, \seq{\omega}{m}{n}=\seq{\omega}{m}{n}}, \quad m \leq n.$$
It differs therefore slightly from our definition (\ref{eg}),(\ref{varm}). The correspondence is $var'_{m}\bra{p(n,.)}=var_{n-m}\bra{p(n,.)}$. The definition of $v(p)$ takes this correspondence into account.
}
%
$$
v(p)= \sup_{m' \in \setZ} \, \sum_{n \geq m'} var_{n-m'}\bra{p(n,.)}.
$$
%
If $m(p) > 0$ and $v(p) < \infty$,
then there exists at most one Gibbs measure consistent with it . 
\ep

So, to prove the uniqueness we only have to establish
that
\beq\label{m(p)sup}
m(p) >0,
\eeq
\beq\label{v(p)sup}
v(p) < + \infty.
\eeq

\bpr
\bigskip

\nid\textbf{$m(p) >0$.} 

Recall that:
$$p(n,\omega)=\prod_{k=1}^N \bra{\omega_k(n) \, \pi\pare{\Xk{n-1,\omega}}
\, + \, \pare{1-\omega_k(n)} \, \pare{1-\pi\pare{\Xk{n-1,\omega}}}}.$$ 
From (\ref{bounds_Vkdet}), (\ref{bounds_sigma_k}) we have:
\beq\label{bounds_X}
-\infty < \frac{\theta - V_k^+}{\sigma_k^+} < \Xk{n-1,\omega}< 
\frac{\theta - V_k^-}{\sigma_k^-} < + \infty.
\eeq
Since $\pi$, given by (\ref{pi}), is monotonously decreasing, we have: 
$$0 < \pi_k^- \deq \pi\pare{\frac{\theta - V_k^-}{\sigma_k^-}}
< \pi\pare{\Xk{n-1,\omega}}< 
\pi_k^+ \deq \pi\pare{\frac{\theta - V_k^+}{\sigma_k^+}} < 1,$$
so that:
\beq\label{boundsPk}
\baR{ccc}
0 < \Pi_k^- \deq \min\pare{\pi_k^-,1-\pi_k^+} < \omega_k(n) \, \pi_k^-
\, + \, \pare{1-\omega_k(n)} \, \pare{1-\pi_k^+}
<p_k(n,\omega)\\
 < 
\omega_k(n) \, \pi_k^+
\, + \, \pare{1-\omega_k(n)} \, \pare{1-\pi_k^-}
< \Pi_k^+ \deq \max\pare{\pi_k^+,1-\pi_k^-} < 1.
\eaR
\eeq%
Finally,
$$
m(p)> \prod_{k=1}^N \Pi_k^- >0,$$
which proves (\ref{m(p)sup}). This also proves the non-nullness of the system of transition probabilities.\\

\bigskip

\nid\textbf{$v(p) < \infty$.} 

The proof, which is rather long, is given in the appendix.

\epr

\su{Consequences} \label{Scons}

\ssu{The probability that  neuron $k$ does not fire in the time interval $[s,t]$.} \label{SPsilent}

In section \ref{SVks} we argued that that this
probability vanishes exponentially fast with $t-s$. 
This probability is $\mu\bra{\bigcap_{n=\ent{s}+1}^{[t]} \Set{\omega_k(n)=0}}$.
 We now prove this result.

\bp
The  probability that neuron $k$ does not fire within the time interval $[s,t]$, $t-s>1$, has
the following bounds:
$$0 < \Pi_-^{\ent{t}-\ent{s}} <
\mu\bra{\bigcap_{n=\ent{s}+1}^{[t]} \Set{\omega_k(n)=0}}
< \Pi_+^{\ent{t}-\ent{s}} < 1,$$
for some constants $0<\Pi_- <\Pi_+ < 1$ depending on 
 the system parameters $\gamma \in \cH$.
\ep

\bpr
We have:
$$  \mu\bra{\bigcap_{n=\ent{s}+1}^{[t]} \Set{\omega_k(n)=0}}=
\int_{\seq{\cA}{-\infty}{\ent{s}}}  
\mu\brac{
\bigcap_{n=\ent{s}+1}^{[t]} \Set{\omega_k(n)=0}}{\sif{\ent{s}}} \, d\mu(\omega)
$$
$$
=
\int_{\seq{\cA}{-\infty}{\ent{s}}}  
\prod_{n=\ent{s}+1}^{[t]}
\mu\brac{\Set{\omega_k(n)=0}}{\sif{n-1}} \, d\mu(\omega)
$$
$$
=\int_{\seq{\cA}{-\infty}{\ent{s}}}  
\prod_{n=\ent{s}+1}^{[t]}
\Probc{\omega_k(n)=0}{\sif{n-1}} \, d\mu(\omega),
$$
where $\Probc{\omega_k(n)=0}{\sif{n-1}}$ is given by
(\ref{Pt+1_cond}) and obeys the bounds (\ref{boundsPk}).
Therefore, setting $\Pi_-= \prod_{k=1}^N \Pi_k^-$ and $\Pi^+= \prod_{k=1}^N \Pi_k^+$,
we have 
$$ \Pi_-^{\ent{t}-\ent{s}} \int_{\seq{\cA}{-\infty}{\ent{s}}} d\mu(\omega)
= \Pi_-^{\ent{t}-\ent{s}}
\leq \mu\bra{\tau_k(t,\omega) \leq s} 
\leq
\Pi_+^{\ent{t}-\ent{s}} \int_{\seq{\cA}{-\infty}{\ent{s}}} d\mu(\omega)
= \Pi_+^{\ent{t}-\ent{s}}.$$
\epr

\ssu{Back to spike trains analysis with the maximal entropy principle}

Here we shortly develop the consequences of our results in relation with the statistical model estimation
discussed in the introduction. A more detailed discussion will be published elsewhere (in preparation and \cite{cessac-palacios:11}).
Set:
\beq\label{phi}
\phi\pare{n,\omega} \deq \log p\pare{n,\omega} = \sum_{k=1}^N \phi_k\pare{n,\omega},
\eeq
 with,
\beq\label{Gibbs_pot}
\phi_k\pare{n,\omega} \deq 
\omega_{k}(n) \, \log \pi\pare{\Xk{n-1,\omega}} \, + \,
\pare{1-\omega_{k}(n)} \,
\log\pare{1-\pi\pare{\Xk{n-1,\omega}}}.
\eeq
The function  $\phi$ is a Gibbs potential \cite{georgii:88}. 
Indeed, we have $\forall m<n$, $\forall \sif{n}$:
%
$$\mu\bra{\seq{\omega}{m}{n} \, | \, \sif{m-1}}= \exp{\sum_{l=m}^n} \phi\pare{l,\omega}.
$$
%
This equation emphasizes the connection with Gibbs distributions in 
 statistical physics which considers probability distributions on multidimensional
lattices with specified boundary conditions and their behavior under space translations \cite{georgii:88}. The correspondence
with our case is that ``time'' is represented by a mono-dimensional space 
and where the ``boundary conditions'' are the past $\sif{m-1}$. Note that in our case the partition function is equal to $1$.

For simplicity assume stationarity (this is equivalent to assuming a time-independent external
current).
 In this case, it is sufficient to consider the potential at time 
$n=0$.

Thanks to the bounds (\ref{bounds_X})
one can make a series expansion of the functions
$\log\pare{\pi}$ and $\log\pare{1-\pi}$ and rewrite the potential under
the form of the expansion:

\beq \label{expansion_phi}
\phi\pare{0,\omega} = \sum_{R=1}^{+\infty} \sum_{r=1}^R 
\sum_{\tiny{\Set{(k_1,n_1), \dots, (k_r,n_r) \in \cP(N,R)}}} \lambda_{(k_1,n_1), \dots, (k_r,n_r)} \, 
\omega_{k_1}(n_1) \dots \omega_{k_r}(n_r)
\eeq 
where $\cP(N,R)$ is the set of non-repeated pairs of integers $(k,N)$ with $k \in \Set{1, \dots, N}$
 and $n \in \Set{-R, \dots, 0}$. We call the product $\omega_{k_1}(n_1) \dots \omega_{k_r}(n_r)$ a
 \textit{monomial}. It is $1$ if and only if neuron $k_1$ fires at time $n_1$, \dots, $k_r$ fires at time $n_r$.
 The $\lambda_{{(k_1,n_1), \dots, (k_r,n_r)}}$'s are explicit functions
of the parameters $\gamma$.
Due to the causal form of the potential,
where the time-$0$ spike, $\omega_k(0)$, is multiplied by a function of the past $\bloc{-\infty}{-1}$,
the polynomial expansion does not contain monomials of the form $\omega_{k_1}(0) \dots \omega_{k_r}(0)$,
$r>1$ (the corresponding coefficient $\lambda$ vanishes).\\

Since the potential has infinite range the expansion (\ref{expansion_phi}) contains infinitely many terms. One can nevertheless
consider truncations to a range $R=D+1$ corresponding to truncating the memory of the process to some memory depth $D$.
Note that although truncations with a memory depth $D$ are approximations,
the distance with the exact potential  converges exponentially fast to $0$
as $D\to +\infty$ thanks to the continuity of the potential, with a decay rate controlled by 
synaptic responses and leak rate.  

The truncated Gibbs potential has the
form:
\beq\label{phiD}
\phi^{(D)}\pare{\bloc{-D}{0}} \, = \, \sum_{l} \lambda_l 
\phi_l\pare{\bloc{-D}{0}};
\eeq
where $l$ stands for $(k_1,n_1), \dots, (k_r,n_r)$ and is an enumeration
of the elements in $\cP(N,D+1)$ and where $\phi_l$ is the corresponding
 monomial. 
Due to the truncation (\ref{phiD}), contrarily to (\ref{phi}),  is not normalized. Its partition function\footnote{
For $D>0$ this is $Z(\bloc{-D}{-1}) \, = \, \sum_{\omega(0)} e^{\phi^{(D)}\pare{\bloc{-D}{0}}} $, 
ensuring that $\frac{e^{\phi^{(D)}\pare{\bloc{-D}{0}}}}{Z(\bloc{-D}{-1})} $
 is a conditional probability
$\Probc{\omega(0)}{\bloc{-D}{-1}}$. Hence it is not a constant but a function of the past $\bloc{-D}{-1}$, in a similar way to statistical physics on lattices where the partition function depends on the boundary conditions. Only in the case $D=0$ (memory less) is this a constant.}  is not equal to $1$
and its computation becomes rapidly intractable as soon as the number of neurons and memory depth increases. 

Clearly, (\ref{phiD}) \textit{ is precisely the form of potential 
which is obtained under the maximal entropy principle, where the $\phi_l$'s are  constraints of type ``neuron $k_1$ is firing at time $n_1$,
neuron $k_2$ is firing at time $n_2$, $\dots$``
 and the $\lambda_l$'s the conjugated Lagrange multipliers}.
 Thus, 
using the maximal entropy principle to characterize spike statistics
in the gIF model by expressing constraints in terms of spike events
(monomials), one can at best find an approximation which can be rather bad, especially if those constraints focus on instantaneous
spike patterns ($D=0$) or short memory patterns. Moreover,
increasing the memory depth to approach better the right statistics
leads to an exponential increase in the number of monomials which
becomes rapidly intractable. Finally, the Lagrange multipliers
$\lambda_l$ are rather difficult to interpret.

On the opposite, the analytic form (\ref{phi})  depends only on a
 finite numbers of parameters ($\gamma$) constraining the neural network dynamics, which have a straightforward interpretations
being physical quantities.  This shows that, at least in gIF model,
 the linear Gibbs potential (\ref{phiD}) obtained from the
 maximal entropy principle is not really appropriate, even for 
empirical/numerical purposes, and that 
 a  form (\ref{phi}) where the infinite
memory $\bloc{-\infty}{-1}$ is replaced by $\bloc{-D}{-1}$ could be more efficient although nonlinear.\\

To finish this section let us discuss the link with  Ising model in light of the present work.
Ising model corresponds to a memory-less case, hence to $D=0$. Since the causal structure
of the Gibbs potential forbids monomials of the form $\omega_{k_1}(0) \dots \omega_{k_r}(0)$,
the $D=0$ expansion of the gIF-Gibbs potential corresponds to a \textit{Bernoulli distribution
where neurons are independent $\phi^{(0)}\pare{0,\omega} = 
\sum_{k=1}^N \lambda_k \,  \omega_{k}(0)$}. The Ising model is therefore irrelevant to
approximate the exact potential of gIF model, if one wants to reproduce spike statistics at
the \textit{minimal discretisation time scale} $\delta$ without considering memory effects.

However, in real data analysis people are usually \textit{binning} data,
with a time windows of width $w \sim 10-20$ ms. Binning consists of recoding the raster
plot with spikes amalgamation. The binned raster $b$ consists of ``spikes''
$b_k(n) \in \Set{0,1}$ where $b_k(n)=1$ if neuron $k$ fired \textit{at least once}
in the time window $[n w, (n+1) w[$. In the expansion (\ref{expansion_phi})
this corresponds to collecting all monomials corresponding to $b_k(n)=1$
in a unique monomial. In this way, the binned potential contains \textit{indeed}
an Ising term $\dots$ that  mixes all spike events occurring within the time interval
 $w$. These events  appear simultaneous because of binning, leading to the Ising pairwise term
$b_{k_1}(0) b_{k_2}(0)$ while events occurring on smaller
time scales are scrambled by this procedure.

The binning effect on Gibbs potential requires however a more detailed  description.
This will be discussed elsewhere. 

\su{Discussion.}\label{SDisc}

To conclude this paper we would like to discuss several
consequences and possible extensions of this work.

\ssu{The spike time discretisation}

In gIF model membrane potential evolves continuously while conductance are updated
with spike occurrence considered as discrete events.
Here we discuss this time discretisation.
Actually, there are two distinct questions.

\sssu{The limit of time-bin tending to $0$} 

This limit would correspond to a case where spike is instantaneous
and modeled by a Dirac distribution. As discussed in \cite{cessac-vieville:08} this limit
raises serious difficulties. To summarize, in real neurons firing occurs within a finite time $\delta$ corresponding to the time of raise and fall for the membrane potential. This involves physico-chemical processes which cannot be instantaneous. 
The time curse of the membrane potential during the spike is described by differential equations, like Hodgkin-Huxley's
\cite{hodgkin-huxley:52}.
Although, the time scale $dt$ appearing in the differential equations has the mathematical meaning of being
arbitrary small, on biophysical grounds this time scale cannot be arbitrary small, otherwise
the Hodgkin-Huxley equations loose their meaning. Indeed, they
correspond to an average over microscopic phenomena such as ionic channels dynamics. In particular,  
their time scale must be sufficiently \textit{large}
to ensure that the description of ionic channels dynamics (opening and closing) in terms of \textit{probabilities} is valid so $dt$ must be
larger than the characteristic time of opening-closing of ionic channels $\tau_P$.
Additionally, Hodgkin-Huxley's equations uses a Markovian approach (master equation) for the dynamics of $h,m,n$ gates.
This requires that the characteristic time $dt$ is quite a bit larger than the characteristic
time of decay for the time correlations between gates activity $\tau_C$.
Summarizing, we must have $0 < \tau_{C},\tau_{P} < dt < \delta$.
 Thus, on biophysical grounds $\delta$ cannot be arbitrary small.

In our case, the $\delta \to 0$ limit is armless however, provided we keep a non zero refractory period, ensuring that only finitely many spikes occur in a finite time interval. 
Taking the limit $\delta \to 0$ without considering a refractory period raises  mathematical problems. One can in principle have uncountably many spikes in a finite
time interval leading to the divergence of physical quantities like
energy. Also,
one can generate nice causal paradoxes \cite{cessac:10a}. Take a loop with two neurons one excitatory and one inhibitory
and assume instantaneous propagation (the $\alpha$ profile is then represented by a Dirac distribution). Then,
depending on the synaptic weights value one can have a situation where neuron $1$ fires
instantaneously, and make instantaneously $2$ firing which prevents instantaneously $1$ from
firing and so on. So taking the limit $\delta \to 0$ as well as $\tau_{refr} \to 0$ induces pathologies not inherent to our approach
but to IF models.

\sssu{Synchronisation for distinct neurons.}

There is a more subtle issue pointed out in \cite{kirst-timme:09}. We do not only discretize
time for each neuron' spikes, we align the spikes emitted by distinct neurons on a discrete time grid,
as an experimental raster does.   As shown in \cite{cessac-vieville:08}
 this induces, in gIF models with a purely deterministic dynamics (no noise and reset to a \textit{constant 
value}), an artificial synchronisation. 
As a consequence the deterministic dynamics of gIF models has generically only stable periodic orbits, 
although periods can be larger than any accessible computational time in a specific region of the parameters
space. Additionally, these periods increase as $\delta$ decreases. 
The addition of noise on dynamics and on the reset value, as we propose in this paper, removes this synchronization effect.

\ssu{Refractory period} \label{SGrammar}

In the definition of the model we have assumed that the refractory period $\tau_{ref}$ was smaller than $1$. The consequence for raster plots is that one can have two consecutive $1$'s in the spike sequence of a neuron. 
The extension to the case where $\tau_{ref}>1$ is straightforward for spike statistics. Having such a refractory period forbids some sequences.
For example if $1 < \tau_{ref} \leq 2$ then all sequences containing two consecutive $1$'s for one neuron ($1, \, 1$) are forbidden. 
If $2 < \tau_{ref} \leq 3$ sequences containing $1, \, *, \, 1$ for a given neuron, 
where $*=0,1$, are forbidden, and so on. More generally, the procedure consisting of forbidding specific (finite) spike blocks is equivalent to introducing a \textit{grammar} in the spike generation. This grammar can be implemented in the Gibbs potential: forbidden sequences have a potential equal to $-\infty$ (resp. a zero probability). In this case, $X=\cA^{\bbbz}$, the set of all possible rasters, becomes a subset
where forbidden sequences have been removed. 

\ssu{Beyond IF models}\label{Sgencond}
Let us now discuss the extension of the present work to more general models
of neurons. First, one characteristic feature of Integrate-and-Fire models is
the reset which has the consequence that the memory of
activity preceding the spike is lost after reset. Although in the deterministic
(noiseless) case this is a simplifying feature allowing for example
 to fully characterize the asymptotic dynamics of (discrete-time) IF models
\cite{cessac:08,cessac-vieville:08}, here it somewhat
renders more complex the analysis. Indeed, it led us to introduce the notion of ``last
reset'' time and, at some point in the proof (see e.g. eq. (\ref{Ineg_f})), obliged
us to consider several situations (e.g. $\tko \geq t-m$ or $\tkop \geq t-m$
versus $\tko < t-m$ and $\tkop < t-m$ in the proof of continuity, section 
\ref{Scont}). On the opposite, considering a model where no such reset 
occur would simply lead us to consider a model where $\tko \to -\infty$, $\forall \omega, \forall k$.
 This case is already considered in our formalism, and actually,
considering that $\tko \to -\infty$, $\forall \omega, \forall k$
simplifies the proofs (for example it eliminates the second term in eq. 
(\ref{Ineg_f})). 

As a matter of fact, the theorems established in this paper
should  therefore also hold without reset. But this requires to replace
the firing condition (\ref{Fire}) by another condition stating what
the membrane potential does during the spike. Although, it could be
possible to propose an ad hoc form for the spike, it would certainly
be more interesting to extend the results here to models where
neurons activity
depends
on additional variables such as adaptation currents, as in the FitzHugh-Nagumo
model \cite{fitzHugh:55,fitzHugh:61,nagumo-etal:62,fitzhugh:69}, or activation-inactivation
variables as in the Hodgkin-Huxley model \cite{hodgkin-huxley:52}.

The present formalism affords an extension toward
such models,  where the neuron fires whenever its membrane potential belongs to a region of the phase space,
which can be delimited by membrane potentials plus additional variables such as adaptation currents or activation-inactivation
variables,  and where the spike is controlled by the global dynamics
of all these variables.  But, while here the firing of a neuron is described by 
the crossing of a fixed threshold, in the FitzHugh-Nagumo
model it is given by the crossing of a separatrix in the plane (voltage-adaptation current), and by a more complex ``frontier'' in the Hodgkin-Huxley model \cite{cronin:87,guckenheimer-oliva:02}.
One difficulty is to precisely define this region. To our knowledge
there is no clear agreement for the Hodgkin-Huxley model (some authors  \cite{guckenheimer-oliva:02} even suggest that the ``spike region'' could have a fractal frontier). The extension toward FitzHugh-Nagumo seems more manageable.

Finally, the most important difficulty toward extending this paper results to more
realistic neural networks is the definition of the synaptic spike response. In IF models the spike is thought as a punctual ``event'' (typically, an ``instantaneous'' pulse)
while the synaptic response is described by a convolution kernel (the $\alpha$-profile). This leads one to consider a somewhat artificial mixed dynamics
where membrane potential evolves continuously while spike are discrete events.
In more realistic models, one would have to consider kinetic equations for
neurotransmitter release, receptor binding and opening of post-synaptic ionic channels \cite{destexhe-mainen-etal:98,ermentrout-terman:10}. Additionally, the consideration of these mechanics deserves a spatially extended modelling of the neuron, with time delays. In this
case, all variables evolve continuously and the statistics of spike trains
would be characterized by the statistics of return times in the ``spike region''.
This statistics is induced by some probability measure in the phase space; a
natural candidate would be the Sinai-Ruelle-Bowen measure \cite{sinai:72,bowen:75,ruelle:78},
for stationary dynamics, or the time-dependent SRB measure for non-stationary
cases, as defined e.g. in \cite{ruelle:99}. These measures
are Gibbs measures as well \cite{keller:98}.
Here, the main mathematical 
property ensuring existence and uniqueness of such a measure
would be uniform hyperbolicity. To our knowledge conditions
ensuring such a  property in networks has not been
 established yet neither for Hodgkin-Huxley's nor for FitzHugh-Nagumo's
models.

\ssu{Synaptic plasticity}\label{Ssynplast}

As the results established in  this paper hold for any synaptic
weight value in $\cH$, they hold as well for networks underlying
synaptic plasticity mechanisms. The effects of a joint evolution
of spikes dynamics, depending on synaptic weights distributions, and
synaptic weights evolution depending on spike dynamics has been
studied in \cite{cessac-rostro-etal:09}. In particular it has been
shown that mechanisms such as Spike-Time Dependent Plasticity
are related to a variational principle for a quantity, the topological
pressure, derived for the thermodynamic formalism of Gibbs distributions.
In the paper  \cite{cessac-rostro-etal:09} the fact that  spike trains
 statistics were given by a Gibbs distribution was a working assumption.
Therefore, the present work establish a firm ground for  
\cite{cessac-rostro-etal:09}. 

\section*{Acknowledgements}
I would like to thank both reviewers for a careful reading of the manuscript, constructive criticism
and helpful comments. I also acknowledge G. Maillard and J.R. Chazottes for invaluable advices and references.
I am also grateful to O. Faugeras and T. Vi\'eville for useful comments.
This work has been supported by the ERC grant Nervi, the ANR grant KEOPS, and the European
grant BrainScales.

\su{Appendix}

Here we establish (\ref{v(p)sup}).  We use the following lemma.

\blem
For a collection $0 \leq a_k,a'_k \leq 1, \, \forall k=1 \dots N$,
we have
\beq\label{ineq1}
\abs{\prod_{k=1}^N a_k-\prod_{k=1}^N a'_k} 
\leq \sum_{k=1}^N \abs{a_k- a'_k},
\eeq
\elem
This lemma is easily proved by recursion.\\

We have, for $n \in \setZ$, $m \geq 0$ 
$$\vm{p(n,.)}=
\sup
\Setc{ 
\abs{\prod_{k=1}^N a_k -
\prod_{k=1}^N a'_k}
}{\omega \eg{m,n} \omega'},
$$
\nid where:
$$a_k=\omega_{k}(n)\pi\pare{\Xk{n-1,\omega}}+
\pare{1-\omega_{k}(n)}\pare{1-\pi\pare{\Xk{n-1,\omega}}},$$
$$a'_k=\omega'_{k}(n)\pi\pare{\Xk{n-1,\omega'}}+
\pare{1-\omega'_{k}(n)}\pare{1-\pi\pare{\Xk{n-1,\omega'}}}.$$

Therefore, using inequality (\ref{ineq1}),

$$\vm{p(n,.)} 
\leq 
\sum_{k=1}^N 
\sup\Setc{ 
\abs{ a_k - a'_k}
}{\omega \eg{m,n} \omega'}.
$$

The condition $\omega \eg{m,n} \omega'$ implies $\omega_k(n)=\omega_k'(n)$
so that:
$$\abs{a_k-a'_k}=
\abs{
\pi\pare{\Xk{n-1,\omega}}
-
\pi\pare{\Xk{n-1,\omega'}}
}
.
$$

We have 
$$\abs{\pi(\Xk{n-1,\omega})-\pi(\Xk{n-1,\omega'})} 
\leq \abs{\Xk{n-1,\omega}-\Xk{n-1,\omega'}} \|\pi'\|_{\infty},$$ 
with $\|\pi'\|_{\infty}=\sup_{x \in \setR} \abs{\pi'(x)}=\frac{1}{\sqrt{2\pi}}$,
so that 
$$
\vm{p(n,.)}
\leq 
\frac{1}{\sqrt{2 \, \pi}} \, \sum_{k=1}^N \vm{\Xk{n-1,.}}.
$$ 
We have now to upper bound $\vm{\Xk{n-1,.}}=
\sup\Setc{ 
\abs{ \Xk{n-1,\omega} - \Xk{n-1,\omega'}}
}{\omega \eg{m,n-1} \omega'}$. We have
$$
\vm{\Xk{n-1,.}}
\leq
\vm{\theta-\Vkdet{\tau_k(n-1,.),n-1,.}} \, 
\sup_{\omega \in X} \frac{1}{\sk{\tau_k(n-1,\omega),n-1,\omega}}
\,$$
$$ + \, 
\sup_{\omega \in X} 
\abs{\theta-\Vkdet{\tau_k(n-1,\omega),n-1,\omega}} \,  \vm{\frac{1}{\sk{\tau_k(n-1,.),n-1,.}}},
$$
with,
$$
\vm{\theta-\Vkdet{\tau_k(n-1,.),n-1,.}} = \vm{\Vkdet{\tau_k(n-1,.),n-1,.}}$$
$$
\leq \vm{\Vksyn{\tau_k(n-1,.),n-1,.}} \, + \, \vm{\Vkext{\tau_k(n-1,.),n-1,.}}
$$
so that, from (\ref{VarVksyn}), (\ref{Var_Vkext}):
\beq\label{Var_theta_m_Vkdet}
\vm{\theta-\Vkdet{\tau_k(n-1,.),n-1,.}}
\preceq 
\sum_{j=1}^N    \,
\cst{A}{kj}{det} \, P_d\pare{\frac{m}{\tau_{kj}}} \, e^{-\frac{m}{\tau_{kj}}}
+
\cst{B}{k}{det} \, e^{-\frac{m}{\tLk}}
\eeq
where:
%
$$\cst{A}{kj}{det}=\cst{A}{kj}{syn} \, + \, \cst{A}{kj}{ext},
$$
%
(see eq. (\ref{Akjsyn}),(\ref{Akjext})) and:
%
$$\cst{B}{k}{det}=\cst{B}{k}{syn} \, + \, \cst{B}{k}{ext},
$$
%
(see eq. (\ref{Bksyn}),(\ref{Bkext})).\\

Moreover, from (\ref{bounds_sigma_k}),
\beq\label{sup_un_sur_sigma_k}
\sup_{\omega \in X} \frac{1}{\sk{\tau_k(n-1,\omega),n-1,\omega}}
\leq
\frac{1}{\sigma_k^-}.
\eeq

From (\ref{bounds_Vkdet}),
\beq\label{sup_theta_m_Vkdet}
\sup_{\omega \in X} 
\abs{\theta-\Vkdet{\tau_k(n-1,\omega),n-1,\omega}}
\leq \max\pare{\abs{\theta-V_k^-},\abs{\theta-V_k^+}}.
\eeq

Finally,
$$
\vm{\frac{1}{\sk{\tau_k(n-1,.),n-1,.}}} \leq
\vm{\sk{\tau_k(n-1,.),n-1,.}} \sup_{\omega \in X}\frac{1}{\skd{\tau_k(n-1,\omega),n-1,\omega}}$$
$$\leq
\frac{1}{\pare{\sigma_k^-}^2} 
\, \vm{\sk{\tau_k(n-1,.),n-1,.}},
$$
while 
$$\vm{\skd{\tau_k(n-1,.),n-1,.}}
%
%
\geq 2 \,\sigma_k^- \, \vm{\sk{\tau_k(n-1,.),n-1,.}},
$$
from (\ref{bounds_sigma_k}), so that:
$$
\vm{\sk{\tau_k(n-1,.),n-1,.}}
\leq \frac{1}{2 \, \sigma_k^-} \, \vm{\skd{\tau_k(n-1,.),n-1,.}},
$$
and, from (\ref{Var_sigma_k}),
\beq\label{Var_unsur_sk}
\vm{\frac{1}{\sk{\tau_k(n-1,.),n-1,.}}}
\preceq
\frac{1}{2 \, \pare{\sigma_k^-}^3}
\bra{ \, \sum_{j'=1}^N \cst{A}{kj'}{\sigma} \, P_d\pare{\frac{m}{\tau_{kj'}}} \, e^{-\frac{m}{\tau_{kj'}}} \, + \,
\cst{C}{k}{\sigma} \, e^{-\frac{2 m} {\tLk}}}
\eeq

Summarizing (\ref{Var_theta_m_Vkdet}), (\ref{sup_un_sur_sigma_k}),
(\ref{sup_theta_m_Vkdet}),(\ref{Var_unsur_sk})
$$
\vm{\Xk{n-1,.}}
\preceq 
\frac{1}{\sigma_k^-} \, \pare{\sum_{j=1}^N    \,
\cst{A}{kj}{det} \, P_d\pare{\frac{m}{\tau_{kj}}} \, e^{-\frac{m}{\tau_{kj}}}
+
\cst{B}{k}{det} \, e^{-\frac{m}{\tLk}}}$$
$$
+ \, 
\max\bra{\abs{\theta-V_k^-},\abs{\theta-V_k^+}}
\, \frac{1 }{\pare{2\, \sigma_k^-}^3}
\bra{\sum_{j'=1}^N  \cst{A}{kj'}{\sigma} \, P_d\pare{\frac{m}{\tau_{kj'}}} \, e^{-\frac{m}{\tau_{kj'}}} \, + \,
\cst{C}{k}{\sigma} \, e^{-\frac{2 m} {\tLk}}}.
$$
Therefore, we have:
%
$$
\vm{\Xk{n-1,.}}
\preceq \sum_{j=1}^N    \,
\cst{A}{kj}{X} \, P_d\pare{\frac{m}{\tau_{kj}}} \, e^{-\frac{m}{\tau_{kj}}}
+
\cst{B}{k}{X} \, e^{-\frac{m}{\tLk}}
+
\cst{C}{k}{X}\, e^{-\frac{2 m}{\tLk}},
$$
%
for constants $\cst{A}{kj}{X},\cst{B}{k}{X},\cst{C}{k}{X}$. \\

As a consequence, 
$$
var_{m}\bra{p(n,.)} \preceq \frac{1}{\sqrt{2 \, \pi}} \,
\sum_{k=1}^N 
\bra{\sum_{j=1}^N    \,
\cst{A}{kj}{X} \, P_d\pare{\frac{m}{\tau_{kj}}} \, e^{-\frac{m}{\tau_{kj}}}
+
\cst{B}{k}{X} \, e^{-\frac{m}{\tLk}}
+
\cst{C}{k}{X}\, e^{-\frac{2 m}{\tLk}}
}.
$$

Therefore, $\sum_{n \geq m'} var_{n-m'}\bra{p(n,.)}$ is bounded from above
by the series 
$$
 \frac{1}{\sqrt{2 \, \pi}}  \, \sum_{k=1}^N 
\bra{\sum_{j=1}^N    \,
\cst{A}{kj}{X} \, \sum_{n \geq m'} P_d\pare{\frac{n-m'}{\tau_{kj}}} \, e^{-\frac{n-m'}{\tau_{kj}}}
+
\cst{B}{k}{X} \, \sum_{n \geq m'} e^{-\frac{n-m'}{\tLk}}
+
\cst{C}{k}{X}\, \sum_{n \geq m'} e^{-\frac{2 (n-m')}{\tLk}}
}$$
$$=
 \frac{1}{\sqrt{2 \, \pi}} \, \sum_{k=1}^N 
\bra{\sum_{j=1}^N    \,
\cst{A}{kj}{X} \, \sum_{l \geq 0} P_d\pare{\frac{l}{\tau_{kj}}} \, e^{-\frac{l}{\tau_{kj}}}
+
\cst{B}{k}{X} \, \sum_{l \geq 0} e^{-\frac{l}{\tLk}}
+
\cst{C}{k}{X}\, \sum_{l \geq 0} e^{-\frac{2 l}{\tLk}}
},
$$
which converges, uniformly in $m'$.
As a consequence, in (\ref{v(p)sup}), $v(p) < + \infty$ and we are done.


{\ifthenelse{\boolean{publ}}{\footnotesize}{\small}
 \bibliographystyle{bmc_article}  
  \bibliography{odyssee,biblio} }     


\ifthenelse{\boolean{publ}}{\end{multicols}}{}



\section*{Figures}
  \subsection*{Figure \ref{Fspike} - Time course of the membrane potential in our model. The blue dashed curve illustrates the shape
of a real spike, but what we model is the red curve. }

\end{bmcformat}
\end{document}